\documentclass[twocolumn,twocolappendix]{aastex701}

\usepackage{amsmath,mathtools,mathrsfs,amssymb,gensymb}
\usepackage{hyperref}
\usepackage{verbatim}
\usepackage{xspace}
\usepackage{graphicx}
\usepackage{longtable}
\usepackage[T1]{fontenc}

\def\sgra{Sgr~A$^*$\xspace}
\def\m87{M87$^*$\xspace}

\defcitealias{M87PaperI}{EHTC~M87~I}
\defcitealias{M87PaperII}{EHTC~M87~II}
\defcitealias{M87PaperIII}{EHTC~M87~III}
\defcitealias{M87PaperIV}{EHTC~M87~IV}
\defcitealias{M87PaperV}{EHTC~M87~V}
\defcitealias{M87PaperVI}{EHTC~M87~VI}
\defcitealias{M87PaperVII}{EHTC~M87~VII}
\defcitealias{M87PaperVIII}{EHTC~M87~VIII}
\defcitealias{M87PaperIX}{EHTC~M87~IX}

\defcitealias{SgrAPaperI}{EHTC~SgrA~I}
\defcitealias{SgrAPaperII}{EHTC~SgrA~II}
\defcitealias{SgrAPaperIII}{EHTC~SgrA~III}
\defcitealias{SgrAPaperIV}{EHTC~SgrA~IV}
\defcitealias{SgrAPaperV}{EHTC~SgrA~V}
\defcitealias{SgrAPaperVI}{EHTC~SgrA~VI}
\defcitealias{SgrAPaperVII}{EHTC~SgrA~VII}
\defcitealias{SgrAPaperVIII}{EHTC~SgrA~VIII}

\begin{document}

\title{Compression-Driven Kinetic Instabilities in Magnetically Arrested Disks}

\correspondingauthor{Vedant Dhruv}
\email{vdhruv2@illinois.edu}

\author[0000-0001-6765-877X]{Vedant Dhruv}
\affiliation{Department of Physics, University of Illinois, 1110 West Green St., Urbana, IL 61801, USA}
\affiliation{Illinois Center for Advanced Study of the Universe, 1110 West Green Street, Urbana, IL 61801, USA}
\email{vdhruv2@illinois.edu}

\author[0000-0002-1227-2754]{Lorenzo Sironi}
\affiliation{Department of Astronomy and Columbia Astrophysics Laboratory, Columbia University, New York, NY 10027, USA}
\affiliation{Center for Computational Astrophysics, Flatiron Institute, 162 Fifth Avenue, New York, NY 10010, USA}
\email{lsironi@astro.columbia.edu}

\author[0000-0002-2685-2434]{Jordy Davelaar}
\affiliation{Department of Astrophysical Sciences, Peyton Hall, Princeton University, Princeton, NJ 08544, USA}
\email{jordydavelaar@gmail.com}

\author[0000-0003-3483-4890]{Aaron Tran}
\affiliation{Department of Physics, University of Wisconsin–Madison, 1150 University Ave, Madison, WI 53706, USA}
\email{atran@physics.wisc.edu}
\begin{abstract}

Event horizon-scale observations of low-luminosity black hole accretion flows favor magnetically arrested disks, characterized by dynamically important magnetic fields ($\beta\lesssim1$, where $\beta$ is the ratio of plasma thermal pressure to magnetic pressure) and a two-temperature transrelativistic plasma. Motivated by plasma conditions in the synchrotron-emitting regions of these models, we perform 2D particle-in-cell simulations of electron-ion plasmas with a realistic mass ratio, subject to continuous compression perpendicular to the mean magnetic field $\boldsymbol{B}_0$. Conservation of particle magnetic moments drives pressure anisotropy $P_{\perp}>P_{\parallel}$, triggering anisotropy-driven instabilities. For ion plasma beta $\beta_{i0}=0.5$ and ion temperature $k_{\text{B}}T_{i0}/m_i c^2=0.05$, the ion pressure anisotropy is regulated by the ion cyclotron instability, while the mirror mode influences the late-time electron anisotropy. Both species develop nonthermal components at high energies, consistent with stochastic acceleration by cyclotron-scale fluctuations. We characterize how the onset and time evolution of the plasma instabilities, as well as the resulting ion and electron anisotropies and energy spectra, vary with $\beta_{i0}$, $k_{\text{B}}T_{i0}/m_i c^2$, electron-to-ion temperature ratio $T_{e0}/T_{i0}$, and the compression rate. Increasing the thermal energy toward relativistic values raises the anisotropy thresholds for all instabilities observed in our simulations, allowing larger anisotropies to develop. For $T_{e0}/T_{i0}<1$, as expected in collisionless two-temperature accretion flows, the growth of mirror and whistler instabilities is delayed or suppressed, leading to increasingly adiabatic evolution of the electrons. Our findings can be used to inform global fluid models of black hole accretion.

\end{abstract}


\section{Introduction}

In hot and dilute plasmas, Coulomb collisions are infrequent, and the associated mean free path can be comparable to or larger than the characteristic length scales of the system. Conservation of the first adiabatic invariant drives a pressure anisotropy relative to the local magnetic field $\boldsymbol{B}$, i.e., the evolution of $P_{\perp}$ decouples from that of $P_{\parallel}$, where $P_{\perp}$ and $P_{\parallel}$ are the plasma pressures perpendicular and parallel to $\boldsymbol{B}$, respectively. Representative low-collisionality environments include the intracluster medium (ICM; \citealp{schekochihin_plasma_physics_galaxy_clusters_2006, kunz_icm_plasma_physics_2022}), the solar wind (see e.g., \citealp{marsch_solar_wind_coulomb_collisions_1983}), and hot accretion flows around low-luminosity active galactic nuclei (LLAGNs; \citealp{mahadevan_quataert_1997, narayan_adaf_review_1998, quataert_riaf_sgra_review_2003}). The latter motivates this work, particularly the systems observed by the Event Horizon Telescope (EHT): Messier 87* (\m87; \citealp{M87PaperI}) and Sagittarius A* (\sgra; \citealp{SgrAPaperI}).

The theoretical interpretation accompanying the release of event horizon–scale images by the EHT points to strongly magnetized flows with coherent, dynamically important magnetic fields \citep{M87PaperV, M87PaperVIII, M87PaperIX, SgrAPaperV, SgrAPaperVIII}. In particular, the spiral pattern of linear polarization in both sources, the relatively large spatially resolved polarization fraction in \sgra, and the powerful jet in \m87 collectively favor a magnetically arrested disk (MAD) configuration \citep{narayan_mad_2003, igumenshchev_2003, tchekhovskoy_efficient_2011, mckinney_general_2012}. In a MAD, poloidal magnetic flux accumulates near the black hole until magnetic pressure becomes comparable to or exceeds the inflowing plasma's ram pressure, intermittently “arresting” accretion. Magnetic reconnection in the midplane of the accretion disk close to the black hole then expels magnetized flux bundles into the disk. These bundles are advected with the flow and dissipate within a few local dynamical times, producing large-amplitude, quasiperiodic fluctuations in the inner disk and jet.

The EHT theoretical analysis employs libraries of general relativistic magnetohydrodynamics (GRMHD) simulations that implicitly treat the plasma as highly collisional and, in the canonical set, assume a single-temperature fluid, i.e., ideal GRMHD (see e.g., \citealp{narayan_jets_2022, dhruv_v3_grmhd_survey_2025}). To carry out radiative transfer and generate synthetic observables---such as polarized images and spectral energy distributions (SEDs)---a simplified prescription ($R-\beta$ model; \citealp{moscibrodzka_general_2016}), motivated by collisionless Alfv\'{e}nic turbulence \citep{quataert_heating_1998,quataert_gruzinov_particle_heating_1999,howes_prescription_2010, kawazura_thermal_2019}, is used to set the electron temperature in terms of the ion temperature and the local plasma $\beta$ ($\beta\equiv P_{\text{gas}}/P_{\text{mag}}$ is the ratio of  plasma thermal pressure to magnetic pressure)\footnote{Other phenomenological prescriptions for the electron-to-ion temperature ratio have also been proposed \citep[e.g., the `critical-beta' model of][]{anantua_critical_beta_2020}, as have prescriptions calibrated directly against particle-in-cell simulations \citep[e.g.,][]{meringolo_turbulence_electron_temperature_prescription_2023}.}. Typically, a relativistic thermal electron distribution function is employed \citep{Wong_2022_patoka}. However, the assumption of a thermal, single-component fluid is formally inconsistent with the collisionless nature of the plasma in the vicinity of LLAGNs \citep{mahadevan_quataert_1997}. Recent global studies that either extend ideal MHD with leading-order kinetic closures for weakly collisional flows \citep{foucart_nonideal_2017,dhruv_egrmhd_variability_2025}, or model the plasma in a fully kinetic manner with first-principles particle-in-cell (PIC) simulations \citep{galishnikova_grpic_2022, vos_grpic_2025, mehlhaff_grpic_torus_accretion_2026}, show that collisionless effects can influence important aspects of the accretion flow, such as angular momentum transport, particle heating, and the resulting observables.



Kinetic effects in accretion disks have been investigated through several complementary frameworks, reflecting the difficulty of bridging kinetic and global scales. Although some global kinetic efforts have appeared, supported by a growing suite of general-relativistic PIC (GRPIC) frameworks \citep{parfrey_grpic_2019, chen_aperture_grpic_2025, galishnikova_entity_grpic_2026, meringolo_fpic_grpic_2026}, they typically rely on simplifying assumptions (e.g., axisymmetry or a reduced ion-to-electron mass ratio $m_i/m_e$), or exclusively target magnetospheric processes \citep{crinquand_reconnection_radiation_2022, mellah_3d_magnetosphere_2023, yuan_pair_production_gaps_bh_magnetospheres_2025}. Consequently, much of what is known about collisionless microphysics in disks comes from \textit{local} simulations that model a patch of the flow. Within this local paradigm, a large body of work considers turbulent, differentially rotating plasmas in the shearing-box framework \citep{hawley_shearing_box_1995}, using kinetic-MHD closures \citep{sharma_kinetic_mri_shearing_box_2006, sharma_electron_2007}, hybrid-kinetic methods \citep{Kunz_2016}, and fully kinetic simulations \citep{riquelme_shearing_pic_mri_2012,hoshino_mri_reconnection_2d_2013,hoshino_amt_mri_3d_2015,bacchini_pic_shearing_box_pair_plasma_2022, bacchini_colliisonless_mri_ntpa_2024,gorbunov_ion_electron_heating_3d_pic_shearing_box_2025}. Collectively, these studies highlight that pressure anisotropy can generate anisotropic viscous stresses that contribute to angular momentum transport and plasma heating. Complementary kinetic treatments of Alfv\'{e}nic turbulence with plasma conditions appropriate to low-luminosity accretion flows, similarly find that a substantial fraction of the injected large-scale energy can be dissipated through anisotropic viscous heating \citep{arzamasskiy_kinetic_turbulence_2023, squire_viscous_heating_turbulence_2023}. Additionally, fully kinetic shearing-sheet and compressing-box studies, where pressure anisotropy is driven by flux freezing and conservation of the particle magnetic moment, have provided insight into the growth and saturation of kinetic instabilities, and how they regulate particle heating and anisotropy \citep{Kunz_2014,riquelme_ic_shearing_pic_2015, sironi_narayan_electron_heating_ic_2015, sironi_electron_heating_2015, Riquelme_2016, riquelme_stochastic_electron_acceleration_2017, Riquelme_2018, ley_stochastic_ion_acceleration_2019}. A common feature across these works is their emphasis on high-$\beta$ ($\beta\gg1$) plasmas (with the exception of \citealt{riquelme_shearing_pic_mri_2012}), which are most relevant to standard and normal evolution disks \citep{narayan_sane_2012, sadowski_sane_2013}, where magnetic fields are weaker and turbulence is stronger.

In this paper, we focus on low-$\beta$ plasmas relevant to MADs. As we discuss below, an important practical advantage of this regime is that, for low-collisionality accretion flows of transrelativistic plasmas (nonrelativistic ions and ultrarelativistic electrons), it allows simulations with a realistic ion-to-electron mass ratio $m_i/m_e=1836$. We carry out local particle-in-cell (PIC) simulations of electron–ion plasma in a compressing-box setup, described in \cite{sironi_narayan_electron_heating_ic_2015}. Compression (expansion) in the directions perpendicular to the mean magnetic field amplifies (reduces) the magnetic field, and conservation of magnetic moment increases (decreases) the perpendicular particle momentum. The resulting distribution therefore develops pressure anisotropy, $P_{\perp}>P_{\parallel}$ ($P_{\perp}<P_{\parallel}$). The growth of pressure anisotropy drives it toward the threshold of plasma instabilities \citep{stix_plasma_waves_1992}. These instabilities tap the free energy stored in the velocity-space anisotropy to generate electromagnetic field fluctuations that pitch-angle scatter the particles, thereby driving the system back toward marginal stability \citep{kennel_petschek_particle_fluxes_1966, gary_what_is_plasma_instability_1992}. Previous low-$\beta$ studies using similar expanding / compressing-box setups have largely focused on nonrelativistic temperatures appropriate for solar-wind applications \citep{liewer_expanding_box_solar_wind_2001, hellinger_parallel_and_oblique_firehose_hybird_simulations_2006, matteini_firehose_solar_wind_hybrid_simulations_2006)}.

The paper is organized as follows. Section \ref{sec:motivation} motivates the parameter space for our PIC simulations by analyzing the emission structure in global weakly collisional MAD models. Section \ref{sec:numerics_simulation_parameters} details the numerical setup and lists input parameters for the fiducial simulation. We present results from the fiducial case in Section \ref{sec:results_fiducial_case}, and in Section \ref{sec:results_parameter_analysis}, we examine how these results change when varying plasma flow conditions (the full list of simulations considered in this work is provided in Appendix \ref{appendix:simulation_parameters}). Finally, Section \ref{sec:summary} summarizes our findings and limitations of this work, discusses its implications for global modeling of black hole accretion systems, and outlines directions for future work.
\section{Weakly collisional black hole accretion simulations}
\label{sec:motivation}

\begin{figure}
\centering
\includegraphics[,width=\linewidth]{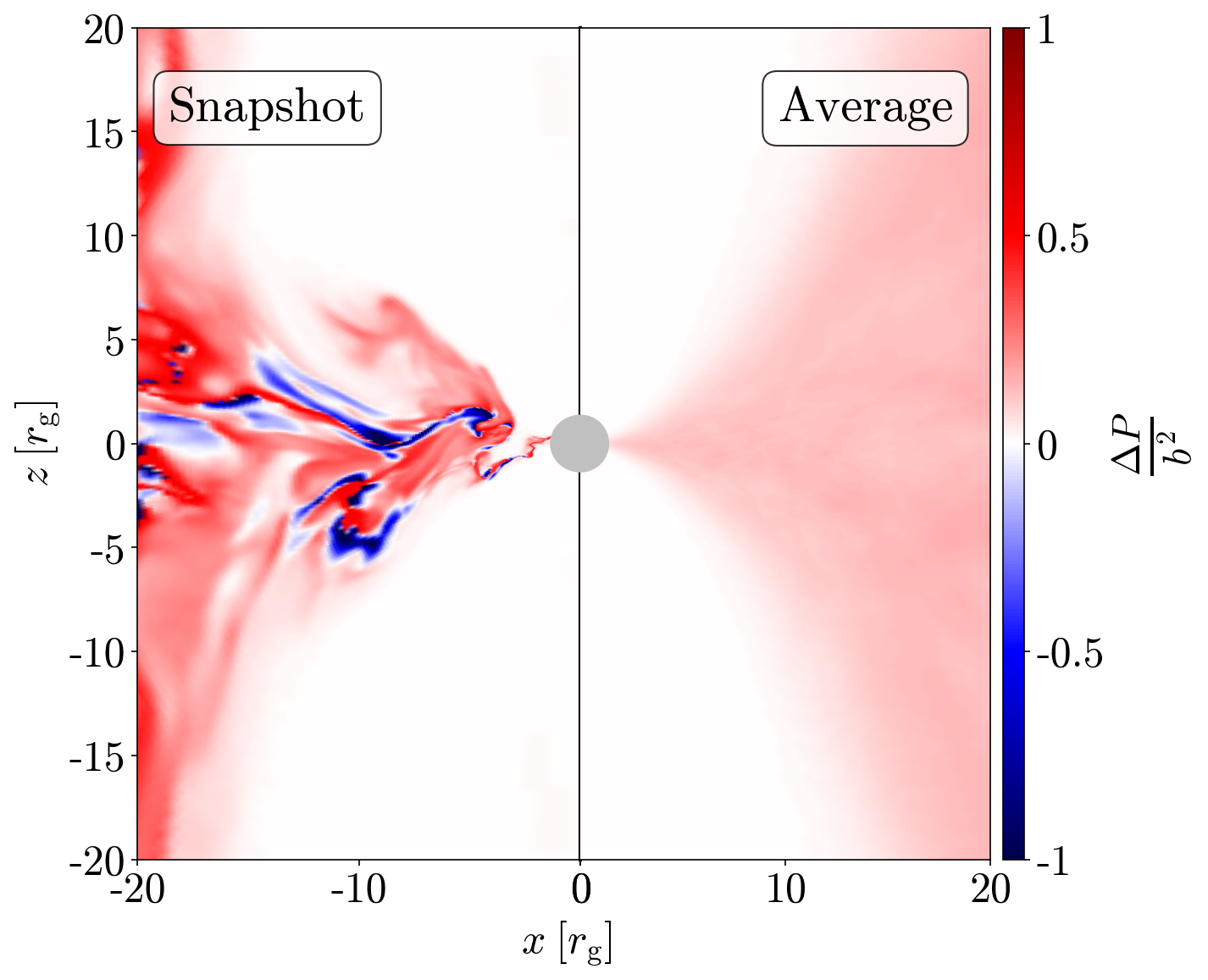}
\caption{Plots of the pressure anisotropy $\Delta P\equiv P_{\perp}-P_{\parallel}$ normalized by the comoving magnetic energy density $b^2$ in the poloidal $(r,\theta)$ plane for a MAD weakly collisional GRMHD simulation from \cite{dhruv_egrmhd_variability_2025}. The gray circle in the center represents the black hole. The left  panel shows a single snapshot at azimuth $\phi=\pi$ while the right panel plots the time- and azimuth-averaged value. We find most of the accretion disk is driven toward $\Delta P>0$.} 
\label{fig:egrmhd_pressure_anisotropy}
\end{figure}

\begin{figure*}
\centering
\includegraphics[width=\textwidth]{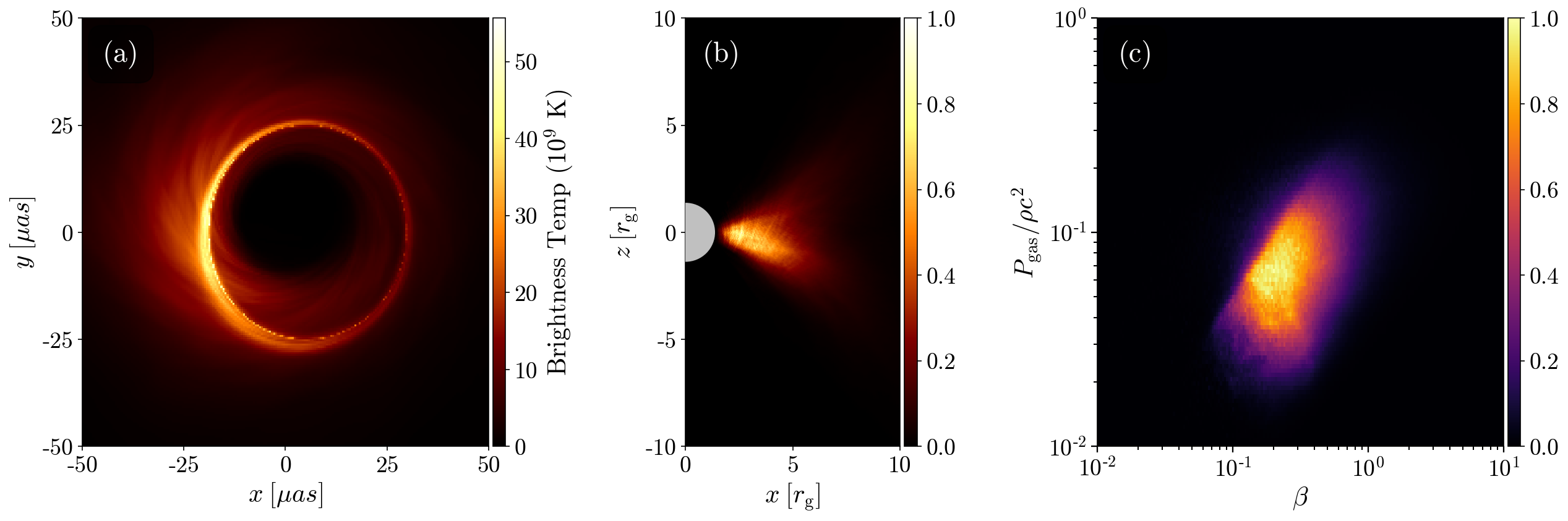}
\caption{Radiative transfer diagnostics for a weakly collisional MAD $a_{*}=+0.94$ simulation ($a_*\equiv Jc/GM^2$ is the dimensionless black hole spin). Left panel: Time-averaged total intensity image at 230 GHz plotted as brightness temperature. Middle panel: Normalized emission map highlighting regions in the simulation domain that contribute to the observed image (left panel). Note that most of the emission for the MAD model originates close to the black hole and near the midplane. Right panel: Emission-weighted distribution of the fluid plasma $\beta$ and temperature $P_{\rm{gas}}/\rho c^2$.}
\label{fig:egrmhd_histogram}
\end{figure*}

In a recent work, \cite{dhruv_egrmhd_variability_2025} simulate accretion onto a supermassive black hole with a weakly collisional plasma model that goes beyond ideal GRMHD by incorporating leading-order kinetic effects arising from long mean free paths along the magnetic field $\boldsymbol{B}$ \citep{chandra_extended_2015}. Specifically, the model includes (i) an anisotropic viscous stress defined with respect to the local magnetic field and (ii) a scalar heat flux acting along the field lines and driven by the field-parallel temperature gradient \citep{braginskii_transport_1965}. The approach is motivated by the following hierarchy of scales $\rho_L\ll r_\text{g}\ll \lambda_{\text{mfp,C}}$, with $\rho_L$ the gyroradius, $r_\text{g}\equiv GM/c^2$ the gravitational radius, and $\lambda_{\text{mfp,C}}$ the Coulomb mean free path. At first glance, this ordering might seem to invalidate a fluid description of low luminosity accretion flows. In practice, pitch-angle scattering driven by kinetic instabilities \citep{sharma_kinetic_mri_shearing_box_2006, Kunz_2014, Riquelme_2016, Riquelme_2018} yields an \emph{effective} mean free path well below that set by Coulomb collisions. This is also supported by observations of the solar wind \citep{coburn_solar_wind_collisionality_2022}.

Figure \ref{fig:egrmhd_pressure_anisotropy} shows the instantaneous (left) and time- and azimuth-averaged (right) pressure anisotropy, $\Delta P \equiv P_{\perp} - P_{\parallel}$, normalized by the comoving magnetic energy density $b^{2}$, for a MAD simulation from \cite{dhruv_egrmhd_variability_2025}. Across the bulk of the disk $P_{\perp} > P_{\parallel}$, as large-scale shear and compressive motions stretch and amplify the field, driving a positive anisotropy. The simulation captures the effects of the mirror and firehose instabilities by imposing pressure anisotropy limiters calibrated to kinetic results \citep{Kunz_2014}: as $\Delta P$ approaches the relevant threshold, the limiters restrict further growth, ensuring the anisotropy remains within the stable regime. A substantial fraction of the disk mass resides near the mirror threshold, $\Delta P_{\mathrm{mirror}}=P_{\parallel}/\beta_{\perp}$, where $\beta_{\perp}\equiv2P_{\perp}/b^2$. The prevalence of $P_{\perp}>P_{\parallel}$ motivates our compressing-box PIC setup.

\cite{dhruv_egrmhd_variability_2025} construct a library of synthetic images and SEDs for the Galactic Center and find that the time-averaged observables of the weakly collisional model closely match those from ideal GRMHD. Figure \ref{fig:egrmhd_histogram} shows time-averaged synchrotron-emission diagnostics from the same MAD weakly collisional simulation as in Figure \ref{fig:egrmhd_pressure_anisotropy}. Because these simulations evolve a single-temperature fluid, the electron temperature is specified during radiative transfer calculations using a prescription that depends on the local $\beta$ \citep{moscibrodzka_general_2016}. In Figure \ref{fig:egrmhd_histogram}, we consider a model where $T_i/T_e\simeq40$ for $\beta\gg1$ and $1.5\lesssim T_i/T_e\lesssim20$ when $0.1\lesssim\beta\lesssim1$. The left panel shows the time-averaged total intensity at the EHT observing frequency, and the middle panel maps the corresponding emission structure in the poloidal plane, highlighting the regions of the flow that dominate the observed output. Most of the emission originates within $r \lesssim 5\,r_{\text{g}}$, where the magnetic pressure is comparable to or exceeds the plasma thermal pressure ($\beta \lesssim 1$). The right panel makes this explicit by showing an emissivity-weighted distribution of the accreting plasma in the $(\beta,\,P_{\mathrm{gas}}/\rho c^2)$ plane, where $P_{\mathrm{gas}}/\rho c^2$ is the dimensionless gas temperature and $\rho$ is the fluid rest-mass density. Emission is concentrated in $0.1 \lesssim \beta \lesssim 2$ and $0.02 \lesssim P_{\mathrm{gas}}/\rho c^2 \lesssim 0.2$. This emissivity-weighted peak defines the parameter space for our PIC study.
\section{Numerical Methods and Setup}
\label{sec:numerics_simulation_parameters}

Large-scale motions such as shear, compression, and expansion modify the magnetic field in collisionless accretion flows, thereby driving pressure anisotropy ($P_{\perp}\neq P_{\parallel}$). Motivated by results of weakly collisional GRMHD simulations (Section \ref{sec:motivation}), in this paper we perform PIC simulations of a compressing box that is meant to represent a local patch in the accretion flow. We consider compression perpendicular to the mean magnetic field, which amplifies the field in response to flux freezing, and as a result increases the perpendicular momenta of the particles due to conservation of the first adiabatic invariant \citep{northrop_adiabatic_charged_particle_motion_1963}. The component of particle momentum parallel to the local magnetic field remains constant. 

We use the relativistic, electromagnetic, three-dimensional PIC code TRISTAN-MP \citep{spitkovsky_tristan_2005, tristan_ascl_2019} and adopt the compressing-box setup described in \cite{sironi_narayan_electron_heating_ic_2015}. There are two reference frames of interest: the laboratory frame $\boldsymbol{x}_L$ and the plasma comoving frame $\boldsymbol{x}$. These frames are related by a Lorentz boost $\boldsymbol{\Lambda}(\boldsymbol{U})$, where $\boldsymbol{U}$ is the relative velocity between the two,
\begin{equation}
    \label{eqn:unprimed_to_lab_frame_transformation}
    dx_L^{\mu} = \Lambda^{\mu}_{\nu}dx^{\nu}.
\end{equation}
To first order in $\vert\boldsymbol{U}\vert/c$, the Lorentz transformation simplifies to
\begin{align}
    \label{eqn:unprimed_to_lab_frame_transformation_first_order}
    dt_L &= dt + \boldsymbol{U}/c^2\cdot d\boldsymbol{x}, \\
    d\boldsymbol{x_L} &= \boldsymbol{U}dt+d\boldsymbol{x}.
\end{align}
The code solves Maxwell's equations and the Lorentz force in a rescaled coordinate system $\boldsymbol{x'}$ also defined in the comoving frame. The rescaled frame is related to $\boldsymbol{x}$ through a time-dependent transformation matrix $\boldsymbol{L}(t)$,
\begin{equation}
\label{eqn:primed_to_unprimed_frame_transformation}
    d\boldsymbol{x} = \boldsymbol{L}d\boldsymbol{x'}.
\end{equation}
For a pure compression or expansion, $\boldsymbol{L}$ is given by
\begin{equation}
    \label{eqn:transformation_matrix}
   \boldsymbol{L}(t) \equiv \frac{\partial\boldsymbol{x}}{\partial\boldsymbol{x'}} = 
   \begin{pmatrix}
    a_{x}(t) & 0 & 0 \\
    0 & a_{y}(t) & 0 \\
    0 & 0 & a_{z}(t)
   \end{pmatrix},
\end{equation}
where $a_x,\,a_y,\,a_z$ are scale factors that quantify the compression or expansion along each of the coordinate axes. Therefore, $\boldsymbol{x'}$ corresponds to a frame where a particle subject only to compression (or expansion) remains at rest, i.e., it has fixed coordinates in that frame. A detailed derivation of the modified evolution equations in the primed coordinate frame is given in \citealp[Appendix A]{sironi_narayan_electron_heating_ic_2015}.

We perform 2D simulations in the $(x,y)$ plane and employ periodic boundary conditions in all directions. We initialize a uniform field $\boldsymbol{B}_0$ along the $y$-axis and apply compression along $x$ and $z$ axes ($a_x,\,a_z<1$ and $a_y=1$). Specifically, the scale factors evolve as $a_x(t) = a_z(t) = (1+qt)^{-1}$, where the compression rate $q$ is a user-provided parameter and is specified in units of the initial ion cyclotron frequency $\omega_{c,i0}\equiv\vert q_e\vert B_0/(\langle\gamma_{i}\rangle m_i c)$, $q_e$ is the electron charge, and $\langle\gamma_i\rangle$ is the average ion Lorentz factor. Hereafter, `$\langle\rangle$' represents an average over the particle distribution function. Compression along the $x$ and $z$ axes results in an increase in the particle number density of each species $n(t)=n_0\,(1+qt)^2$ due to conservation of the total number of particles, and an increase in the mean field strength $B^y(t)=B_0\,(1+qt)^2$ due to flux freezing.

To isolate the role of ion-only and electron-only physics,  we perform additional simulations in which one of the species undergoes uniform, isotropic compression. Then, this species does not develop pressure anisotropy, and primarily serves to maintain charge neutrality. We achieve this by modifying the particle pusher so that the compression matrix $\boldsymbol{L}$ for that species corresponds to $a_x=a_y=a_z=a_{\text{iso}}$, where $a_{\text{iso}}=(1+q_{\text{iso}}t)^{-1}$ and $q_{\text{iso}}=2q/3$ \citep{tran_ic_icm_2023}. We also perform 1D simulations---including isotropic-ions and isotropic-electrons control runs---to isolate the role of waves propagating parallel to the background magnetic field, such as ion and electron cyclotron modes, since this setup naturally suppresses oblique mirror instabilities. 

Motivated by Figure \ref{fig:egrmhd_histogram}(c), our fiducial PIC simulation employs an initial ion plasma beta $\beta_{i0}=0.5$ and an initial ion dimensionless temperature $\Theta_{i0}\equiv k_{\text{B}}T_{i0}/m_i c^2=0.05$. We focus on the transrelativistic regime ($\Theta_e\equiv k_{\text{B}}T_{e}/m_e c^2\gg1$), where the relativistic inertia of the electrons reduces the scale separation between species. The resulting increase in the effective electron mass ($\langle\gamma_e\rangle m_e \lesssim m_i$, where $\langle\gamma_e\rangle$ is the average electron Lorentz factor), enables the use of a realistic ion-to-electron mass ratio $m_i/m_e = 1836$. For the fiducial simulation we choose $T_{e0}/T_{i0}=1$, but study how these results are affected if we consider an initially colder population of electrons, i.e., $T_{e0}<T_{i0}$, in Section \ref{sec:results_electron_to_ion_temperature_ratio}. This is motivated by models of hot, dilute accretion flows in which inefficient ion–electron coupling leads naturally to a two-temperature plasma \citep{shapiro_1976, rees_hot_tori_1982, mahadevan_quataert_1997}. It is also supported by recent horizon-scale EHT observations of the Galactic Center and M87* \citep{M87PaperV, SgrAPaperV}, which favor GRMHD models with $T_e<T_i$. The ion cyclotron frequency close to the black hole ($r\sim 5\,GM/c^2$) is expected to exceed the local dynamical frequency, $\Omega_{\rm dyn}\sim \sqrt{GM/r^3}$, by several orders of magnitude, typically $\sim 10^{7}$--$10^{8}$. To keep the simulations tractable we consider $\omega_{c,i0}/q=1600$ in the fiducial case, but discuss implications of slower compression rates in Section \ref{sec:results_compression_rate}. Appendix \ref{appendix:simulation_parameters} lists all the simulations considered in this work along with their corresponding input parameters.

For the fiducial simulation, we set the initial electron skin depth $c/\omega_{p,e0} = 10$ and the overall box size $L_x = L_y = 880$ cells, where $\omega_{p,e0}\equiv\sqrt{4\pi n_0 q_e^2 / (\langle\gamma_{e}\rangle m_e)}$ is the initial electron plasma frequency. This ensures that the simulation domain spans at least 30 times the longest characteristic length scale in the system, providing sufficient resolution to capture both ion- and electron-scale instabilities. For example, the initial ion gyroradius $\rho_{L,i0} = \langle\gamma_i v_{\perp, i0}\rangle m_i c / (\vert q_e\vert B_0)$---where $v_{\perp,i0}$ is the initial ion perpendicular velocity---spans $\sim21$ cells, and the ion skin depth $c/\omega_{p,i0}$ is resolved with $\sim26$ cells. By comparison, the initial electron gyroradius $\rho_{L,e0}$ spans 11 cells, underscoring that the ion–electron kinetic scale separation is substantially reduced in the transrelativistic regime, even for realistic mass ratios. In Appendix \ref{appendix:scales}, we express the characteristic length and frequency scales for both species in terms of the simulation input parameters, accounting for their relativistic inertia. To ensure numerical stability throughout the simulation, including during the late stages of compression ($qt \simeq 2$), we set the speed of light in code units to 0.13 cells per timestep \citep[Appendix A]{sironi_narayan_electron_heating_ic_2015}. We utilize 1024 particles per cell (\texttt{ppc}; 512 per species) for the fiducial case, and demonstrate numerical convergence with respect to particle number in Appendix \ref{appendix:convergence}.

\section{Plasma Instabilities in a Compressing Transrelativistic Plasma}
\label{sec:results_fiducial_case}

\begin{figure}
\centering
\includegraphics[,width=\linewidth]{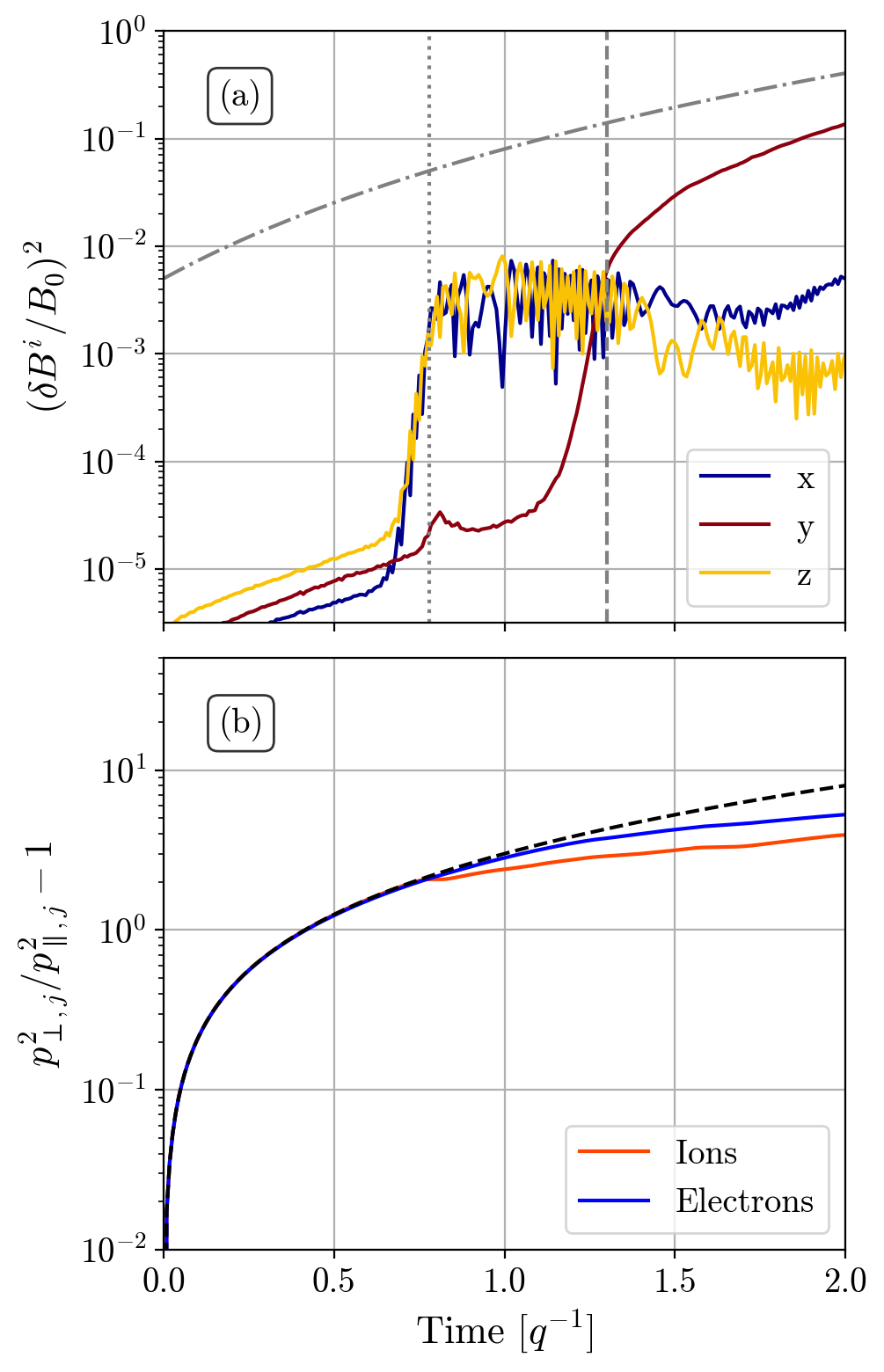}
\caption{Time series of the box-averaged squared magnetic field perturbations in panel (a) and particle anisotropy in panel (b) for the fiducial simulation with $T_{e0}/T_{i0}=1$, $\beta_{i0}=0.5$, and $\Theta_{i0}=0.05$. The gray dash-dot line in panel (a) is $\propto(1+qt)^4$ and is the expected evolution of the $\vert\boldsymbol{B}_0\vert^2$ due to flux-freezing. The black dashed line in panel (b) follows $(1+qt)^2-1$ and is the expected growth rate if the particles underwent pure adiabatic compression. The gray vertical lines in panel (a) mark the snapshot times shown in Figure \ref{fig:fiducial_sim_snapshots}: the dotted line denotes $t=0.78\,q^{-1}$ (top row) and the dashed line denotes $t=1.30\,q^{-1}$ (bottom row).}
\label{fig:fiducial_sim_timeseries}
\end{figure}

\begin{figure*}
\centering
\includegraphics[,width=\textwidth]{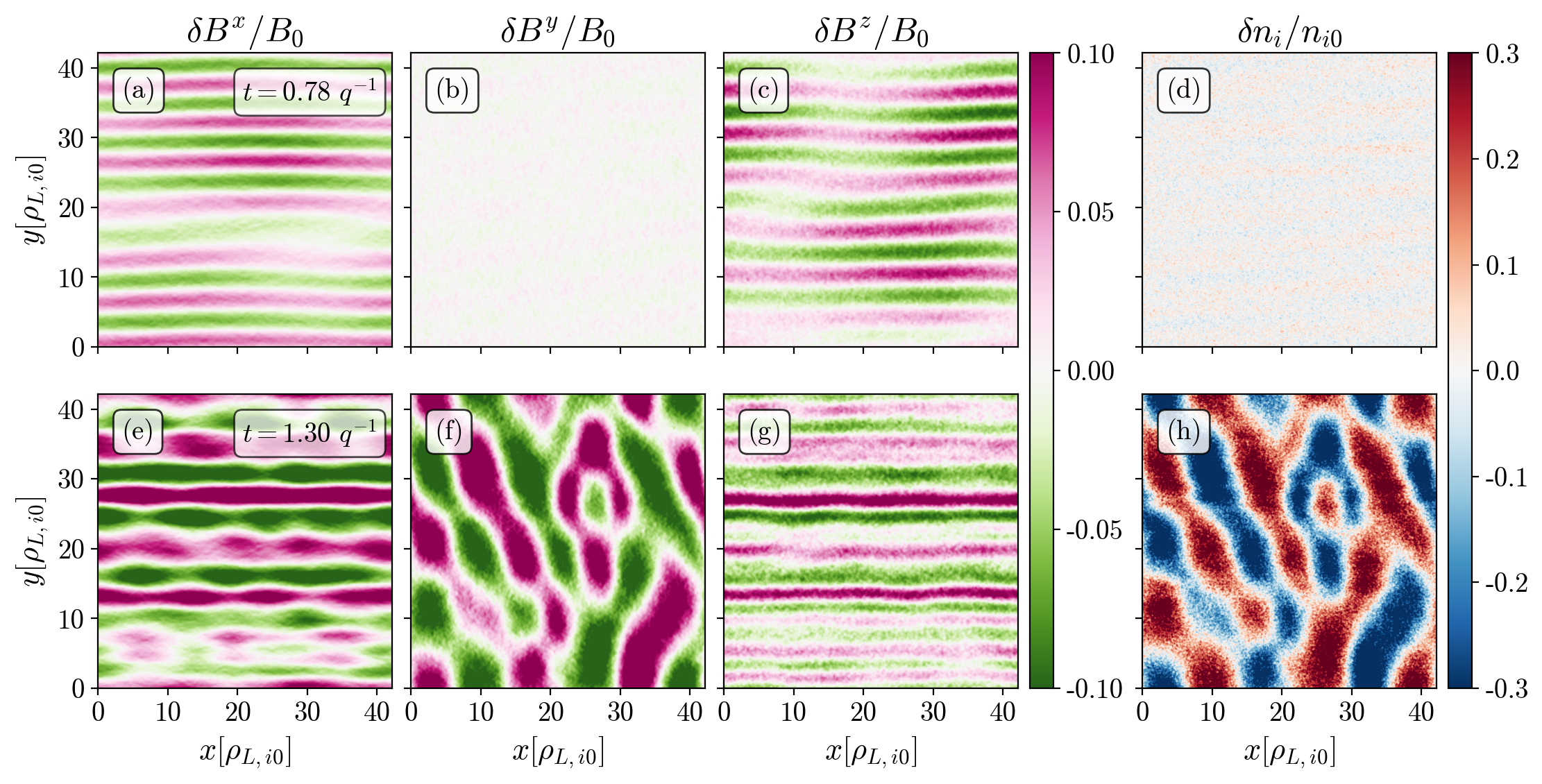}
\caption{Snapshots for fiducial simulation with $T_{e0}/T_{i0}=1$, $\Theta_{i0}=0.05$, $\beta_{i0}=0.5$. Left to right: the perturbations in the x (first column), y (second column) and z (third column) component of the magnetic field normalized by the mean field, and the perturbations in the ion number density normalized by the initial ion number density (fourth column) at two timestamps. The first row shows results at $t=0.78\,q^{-1}$, and the second row shows the same quantities at a later time $t=1.30\,q^{-1}$ once the mirror mode is activated. This is evident by the spatial anticorrelation in the field and number density perturbations.}
\label{fig:fiducial_sim_snapshots}
\end{figure*}

\begin{figure*}
\centering
\includegraphics[width=\textwidth]{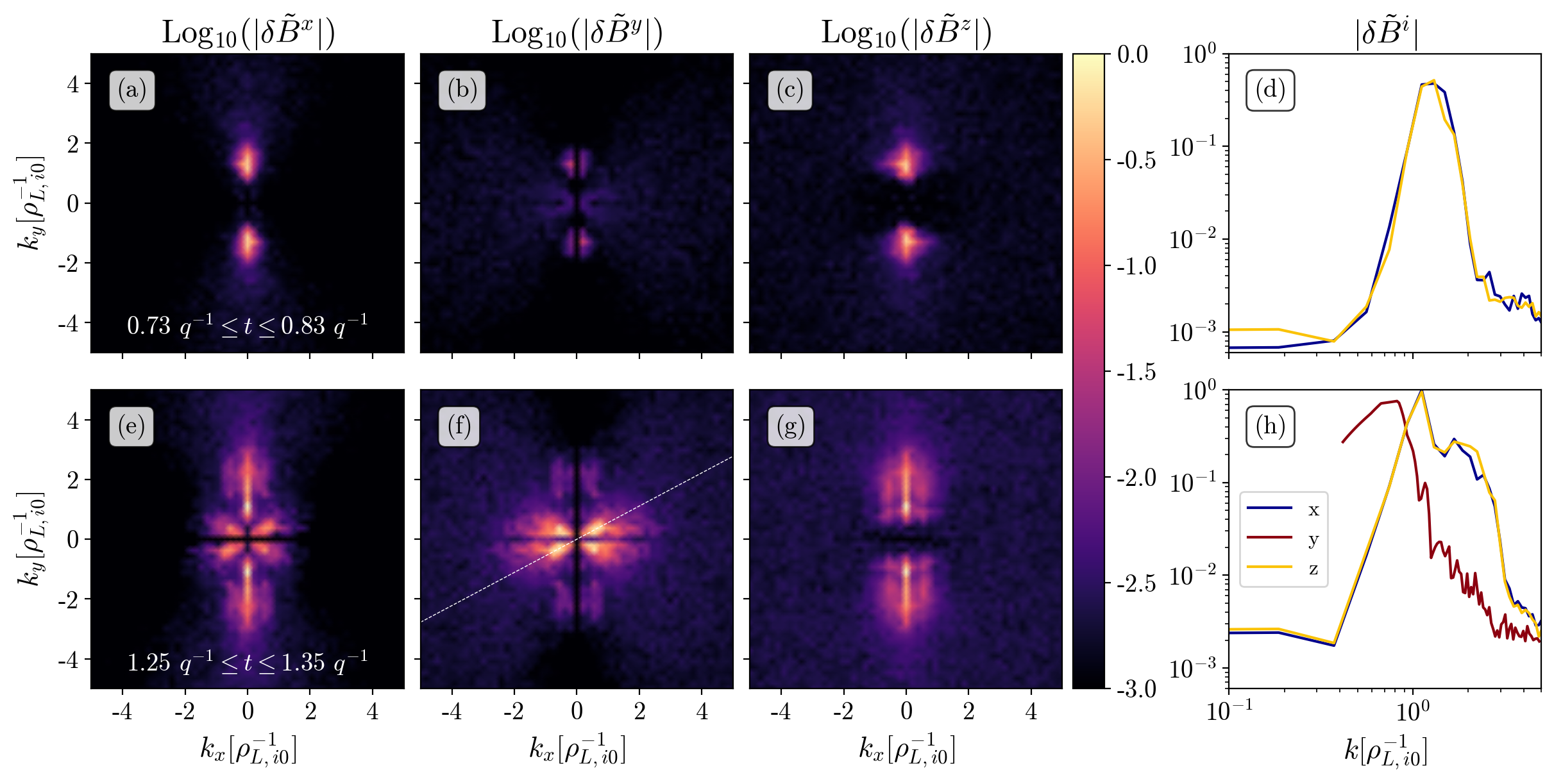}
\caption{Spatial power spectrum of the magnetic field components for the fiducial simulation plotted at two different time intervals. The top row corresponds to $0.73\,q^{-1} \leq t \leq 0.83\,q^{-1}$ and the bottom row to $1.25\,q^{-1} \leq t \leq 1.35\,q^{-1}$. The first three columns plot the logarithm of the 2D power spectrum $\vert\delta\tilde{B}^{i}\,(k_x,ky)\vert$, and the rightmost column shows a 1D slice of $\delta\tilde{B}^{x}$ and $\delta\tilde{B}^z$ along $k_y$. For the time interval $1.25\,q^{-1} \leq t \leq 1.35\,q^{-1}$, in panel (h) we also plot a 1D slice of $\delta\tilde{B}^{y}$ along the wave vector of the oblique mirror mode as measured over that time range (indicated by the dotted white line in panel (f)). The field perturbation in each panel is normalized by its corresponding maximum amplitude.}
\label{fig:fiducial_sim_kspace}
\end{figure*}

This section presents results for the fiducial simulation, initialized with $T_{e0}/T_{i0}=1$, $\beta_{i0}=0.5$, $\Theta_{i0}=0.05$, and $\omega_{c,i0}/q=1600$. Compression perpendicular to $\boldsymbol{B}_0$ drives anisotropy $P_{\perp}>P_{\parallel}$ due to conservation of adiabatic invariants. This is reflected in the early time evolution of the box-averaged particle anisotropy, $A_{j}\equiv p_{\perp,j}^2/p_{\parallel,j}^2-1$ (with $j\in\{i,e\}$ denoting ions or electrons), shown in Figure \ref{fig:fiducial_sim_timeseries}(b), where both ion and electron anisotropies closely track the adiabatic evolution (dashed black line). Throughout this work we quantify anisotropy using the mean squared particle momenta $p_{j}^2\equiv m_j^2\langle(\gamma_j v_j)^2\rangle$ rather than the temperature $k_{\rm{B}}T_j=m_j\langle\gamma_j v_j^2\rangle/3$ as the momentum-based anisotropy is the one naturally tied to the relativistic adiabatic invariants \citep{sturrock_plasma_physics_1994}. Figure \ref{fig:fiducial_sim_timeseries}(a) shows the time evolution of the box-averaged, squared magnetic field fluctuations, normalized by the mean field $(\delta B^i /B_0)^2$. The slow rise in the field fluctuations prior to the onset of the instabilities (until $t\simeq0.65\,q^{-1}$) reflects amplification of magnetic fluctuations seeded by particle noise \citep{birdsall_langdon_pic_1991, melzani_apart_2013}.

At $t\simeq0.65\,q^{-1}$, the ion anisotropy exceeds the ion cyclotron instability threshold, triggering exponential growth of transverse magnetic fluctuations ($x$ and $z$ components) as ions with large pitch angles transfer energy to the electromagnetic fields. This exponential phase continues till $t\simeq0.8\,q^{-1}$, when the amplified magnetic fields become strong enough to pitch-angle scatter the ions and limit the anisotropy at its marginally stable value \citep{kennel_petschek_particle_fluxes_1966}. The top row in Figure \ref{fig:fiducial_sim_snapshots} shows the spatial distribution---in the $(x,y)$ plane of the simulations---of the field and density perturbations toward the end of the exponential phase of the ion cyclotron instability. The presence of transverse field perturbations propagating predominantly along the magnetic field is characteristic of the ion cyclotron mode. Linear theory predicts that this mode exhibits its maximum growth rate when $\boldsymbol{k}\times\boldsymbol{B}_0=0$, where $\boldsymbol{k}$ is the wave vector \citep{davidson_ogden_ic_instability_1975, gary_anisotropy_instabilities_solar_wind_1976, stix_plasma_waves_1992}. The ion cyclotron instability is a resonant instability, interacting with ions that satisfy the gyroresonance condition $\omega-kv_{\parallel, i} = \omega_{c,i}$, where $\omega$ is the wave angular frequency, $k$ is the wavenumber, $v_{\parallel, i}$ is the ion velocity parallel to $\boldsymbol{B}_0$, and $\omega_{c,i}$ is the relativistic ion cyclotron frequency. Linear kinetic theory indicates that hot and anisotropic electrons can affect both the growth rate and the anisotropy threshold of the ion cyclotron instability \citep{kennel_scarf_instabilities_solar_Wind_1968, shaaban_emic_electrons_solar_wind_2015, shaaban_instabilities_solar_wind_2017, wang_emic_electrons_magnetosphere_2023}. We discuss this later in the section by comparing the fiducial simulation with an equivalent simulation where electrons are compressed isotropically, thereby isolating the effect of electron anisotropy.

Resonant pitch-angle scattering of ions by high-frequency electromagnetic fluctuations limits the ion temperature anisotropy, $A_{\text{T,i}}\equiv T_{\perp,i}/T_{\parallel,i}-1$. This upper bound, which sets the marginal stability threshold, is well described by an inverse-$\beta_{\parallel,i}$ scaling, $A_{\text{T,i}} = S_i\,\beta_{\parallel,i}^{-\alpha_i}$, motivated by magnetosheath observations \citep{phan_ic_inverse_beta_magnetosheath_1994, anderson_ic_inverse_beta_magnetosheath_1994, fuselier_ic_inverse_beta_magnetosheath_1994} and by theoretical and numerical calculations \citep{gary_lee_ic_inverse_beta_derivation_1994, gary_ic_inverse_beta_hybrid_simulations_1994}. Here, $0.4\lesssim \alpha_i \lesssim 0.5$ for nonrelativistic ions and $S_i$ is of order unity---typically obtained by fitting to the result of kinetic simulations. For the fiducial case discussed here, the late time temperature anisotropy ($2\,q^{-1}\leq t\leq 3\,q^{-1}$) follows the scaling  $0.45\,\beta_{\parallel,i}^{-0.75}$. Once the anisotropy is restricted, it still rises, but at a slower rate than the initial adiabatic growth, because $\beta_{\parallel ,i}$ continues to drop. This eventually drives the mirror instability at $t\simeq1.1\,q^{-1}$, indicated by the exponential increase in the longitudinal field perturbations\footnote{The minor increase in $\delta B^y/B_0$ near the conclusion of the ion cyclotron instability's exponential growth ($t\sim0.8\,q^{-1}$) is a consequence of the ion cyclotron modes not being perfectly parallel to the mean field, as discussed further in Appendix \ref{appendix:dby_bump}.} $(\delta B^{y}/B_0)^2$. The second row in Figure \ref{fig:fiducial_sim_snapshots} presents magnetic field and ion density perturbations toward the end of the linear phase of the mirror mode (see vertical dashed line in Figure \ref{fig:fiducial_sim_timeseries}(a)). As is characteristic of the mirror instability, the wave vector is oblique to the background magnetic field (Figure \ref{fig:fiducial_sim_snapshots}(f)), and the mode is compressive and non-propagating \citep{chandrasekhar_pinch_stability_1958,vedenov_plasma_stability_1961,tajiri_waves_collisionless_plasmas_kinetic_approach_1967, hasegawa_mirror_instability_1969,southwood_kivelson_mirror_partI_1993, kivelson_southwood_mirror_partII_1996}. As a result of pressure balance, particles bunch up in regions of low magnetic field strength, known as ``magnetic wells''. Consequently, the ion number density perturbations $\delta n_i/n_{i0}$ are anticorrelated with $\delta B^y/B_0$, as evidenced in Figure \ref{fig:fiducial_sim_snapshots}(h).

\begin{figure*}
\centering
\includegraphics[width=\textwidth]{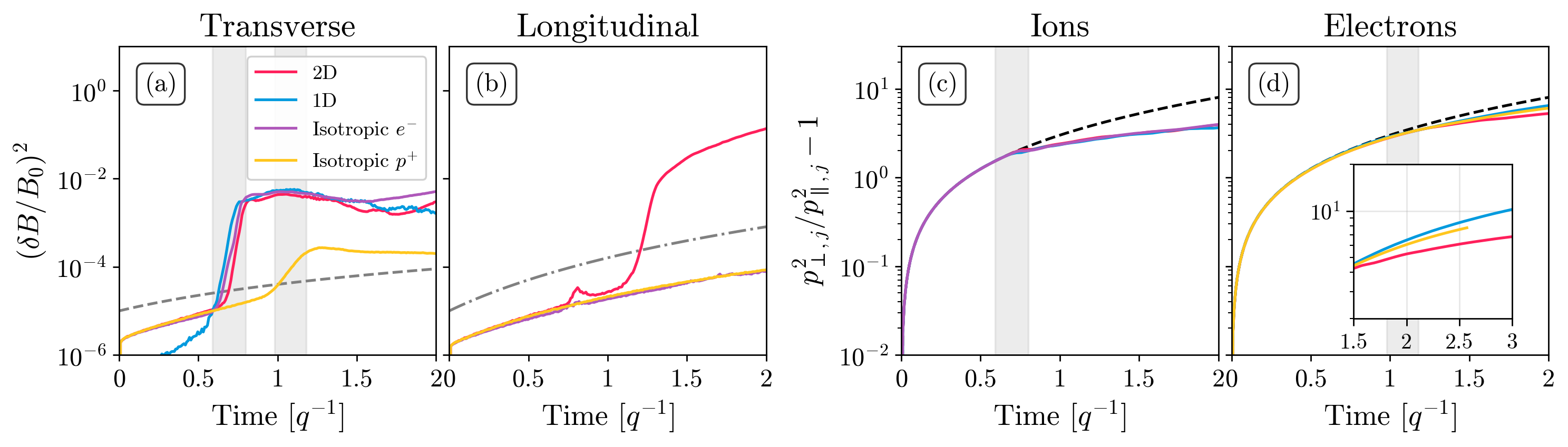}
\caption{A comparison of the fiducial 2D simulation (red), the corresponding 1D simulation (blue), an equivalent 2D simulation where electrons are compressed isotropically (purple), and a 2D simulation where ions are compressed isotropically (yellow). Panel (a) shows fluctuations in the perpendicular magnetic field component, $(\delta B_{\perp}/B_0)^2$. Panel (b) shows fluctuations in the parallel magnetic field component, $(\delta B^y/B_0)^2$. Panel (c) shows the ion anisotropy $A_i$. Panel (d) shows the electron anisotropy $A_e$. The inset in panel (d) shows the electron anisotropy evolution over a longer duration (till $t=3\,q^{-1}$). Vertical gray bands mark the time intervals analyzed in the dispersion plot (Figure \ref{fig:fiducial_dispersion_relation}). The gray dashed (panel (a)) and dot-dashed (panel (b)) lines show $(1+qt)^2$ and $(1+qt)^4$, respectively. The black dashed lines in the right two panels indicate the anisotropy growth expected from purely adiabatic compression.}
\label{fig:fiducial_isotropic_1d_perturbations_anisotropy_comparison}
\end{figure*}

To characterize the spectral properties of the anisotropy-driven waves, we plot the 2D spatial Fourier transform of the magnetic field perturbations $\delta\tilde{B}^i$ at two times in Figure \ref{fig:fiducial_sim_kspace}. These correspond to the two snapshots shown in Figure \ref{fig:fiducial_sim_snapshots}, and mark the end of the exponential growth phases of the ion cyclotron and mirror instabilities, respectively. Each row shows the spectrum averaged over a $0.1\,q^{-1}$ interval centered on the corresponding snapshot time. In the first interval (top row), nearly all power resides in modes parallel to the mean field with a characteristic wavenumber $k_{\parallel}\rho_{L,i0}\sim1$. This is most clearly seen in Figure \ref{fig:fiducial_sim_kspace}(d), which plots $\vert\delta\tilde{B}^x\vert$ and $\vert\delta\tilde{B}^z\vert$ along the mean field (for $k_y>0$) at $k_x=0$. In the second row, we observe that the in-plane magnetic field perturbations possess significant power at an oblique angle relative to the mean field, a key property of mirror modes. The dashed white line denotes the angle $\theta_{\text{mir}}=59.3\degree$ where the power in the longitudinal component is maximized in the $(k_x,k_y)$ plane. We determine $\theta_{\text{mir}}$ by calculating the power-weighted average wavenumbers $\bar{k}_x$ and $\bar{k}_y$ in the first quadrant and evaluating $\theta_{\text{mir}} = \text{tan}^{-1}(\bar{k}_y/\bar{k}_x)$. Figure \ref{fig:fiducial_sim_kspace}(h)---which includes $\delta\tilde{B}^y$ along the dashed white line in Figure \ref{fig:fiducial_sim_kspace}(f) alongside $\delta\tilde{B}^x$ and $\delta\tilde{B}^z$ at $k_x=0$ (as in panel d)---confirms that the fastest growing mirror mode has a wavenumber $k\rho_{L,i0}\lesssim1$.

The anisotropic electrons provide a source of free energy that drives the whistler instability \citep{sudan_whistler_1963, sudan_whistler_1965, gladd_relativistic_whistler_1983} once $A_{\text{T},e}$ exceeds the marginal stability threshold. The onset of the whistler modes coincides with the electron population’s departure from adiabatic growth at $t\simeq1.2\,q^{-1}$ in Figure \ref{fig:fiducial_sim_timeseries}(b). The resulting electromagnetic fluctuations, which maintain the anisotropy at marginal stability via pitch-angle diffusion of resonant electrons, are significantly smaller in magnitude than those generated by ion cyclotron waves. 

To isolate the respective contributions of ion and electron dynamics to anisotropy-driven instabilities, we perform two auxiliary simulations where one species is compressed isotropically, thereby suppressing the growth of its anisotropy. We also conduct a 1D simulation equivalent to the fiducial case, in which both species undergo compression perpendicular to the mean field. By construction, the 1D geometry prevents the growth of the oblique mirror mode, allowing us to quantify the specific role of the mirror instability in regulating plasma anisotropy. 

Figure \ref{fig:fiducial_isotropic_1d_perturbations_anisotropy_comparison} compares the temporal evolution of these different runs. At early times ($t\lesssim0.5\,q^{-1}$), before the instabilities develop, $(\delta B_{\perp}/B_0)^2$ is smaller in the 1D run than in the 2D runs, reflecting its lower noise level due to the much larger particle count (\texttt{ppc}=32768). We find that the properties of the ion cyclotron instability---including the onset time, the linear growth phase, the saturated value of $(\delta B_{\perp}/B_0)^2$, and the subsequent secular evolution (up to the point of mirror mode activation in the fiducial 2D case)---remain unchanged across the fiducial, 1D, and isotropic-electrons simulations. This consistency is illustrated by the nearly perfect overlap of the red, blue, and purple lines in Figure \ref{fig:fiducial_isotropic_1d_perturbations_anisotropy_comparison}(a). We note that $(\delta B_{\perp}/B_0)^2$ includes contributions from $(\delta B^x/B_0)^2$, which can be dominated by mirror modes when mirror fluctuations reach large amplitudes. The absence of significant growth in the parallel field perturbations for the isotropic-electrons simulation, shown in Figure \ref{fig:fiducial_isotropic_1d_perturbations_anisotropy_comparison}(b), demonstrates that the mirror instability is suppressed. This highlights the importance of electron anisotropy in driving the growth of the mirror mode, which is consistent with its nonresonant nature. Previous analytical studies \citep{pokhotelov_electron_anisotropy_mirror_2000} and numerical simulations \citep{remya_electron_anisotropy_ic_mirror_2013} have demonstrated that introducing a net anisotropy in electrons not only increases the mirror growth rate and the range of unstable wavenumbers but also lowers the threshold for instability onset. The growth of transverse perturbations in the isotropic-ions simulation (yellow curve in Figure \ref{fig:fiducial_isotropic_1d_perturbations_anisotropy_comparison}(a)) at $t\simeq1\,q^{-1}$ indicates the onset of whistler waves. Their amplitude is much smaller than the transverse magnetic fluctuations seen in the fiducial 2D and isotropic-electrons simulations, underscoring the subdominant contribution of the whistler instability to the magnetic fluctuations in the fiducial 2D case.

Figure \ref{fig:fiducial_isotropic_1d_perturbations_anisotropy_comparison}(c) compares the ion anisotropy in the 2D fiducial run, the 1D run, and the isotropic-electrons run. The mirror mode appears in the 2D fiducial case, yet $A_i$ evolves nearly identically across all three runs during the nonlinear phase. This indicates that $A_i$ is primarily regulated by the ion cyclotron instability, with the mirror mode playing a subdominant role. The electron anisotropy $A_e$ is, however, affected by the formation of magnetic mirrors, as evidenced by the extended evolution of $A_e$ up to $t=3\,q^{-1}$, shown in the inset of Figure \ref{fig:fiducial_isotropic_1d_perturbations_anisotropy_comparison}(d). Specifically, $A_e$ is lower in the fiducial case relative to the 1D and isotropic-ions simulations. This suggests that the electron anisotropy saturates at a level below the relativistic whistler marginal stability threshold, likely as a result of mirror fluctuations. 

\begin{figure}
\centering
\includegraphics[,width=\linewidth]{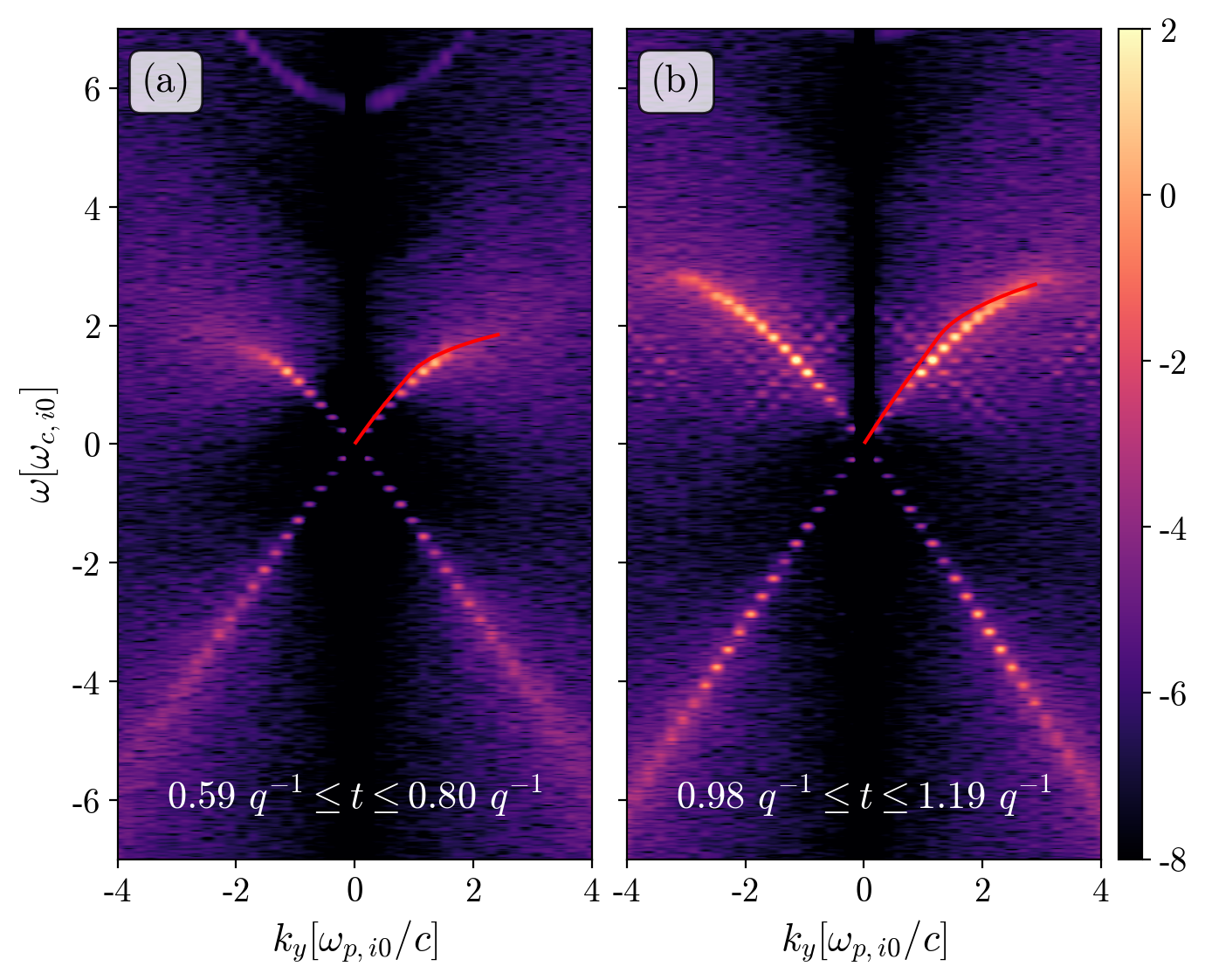}
\caption{Spatiotemporal power spectrum of the magnetic field fluctuations $\delta B^z + i\,\delta B^x$ for the 2D fiducial case. The panels show the dispersion relation $\omega(k_y)$ during the linear growth phase of the ion cyclotron instability (panel (a)) and the whistler instability (panel (b)); see Figure \ref{fig:fiducial_isotropic_1d_perturbations_anisotropy_comparison} for the corresponding growth in field fluctuations and evolution of particle anisotropy. The ion cyclotron modes are LCP ($\omega>0$), while the whistler modes are RCP ($\omega<0$). Overlaid red curves indicate the theoretical dispersion relations for the ion cyclotron mode derived from a nonrelativistic, parallel-propagating linear solver.}
\label{fig:fiducial_dispersion_relation}
\end{figure}

Figure \ref{fig:fiducial_dispersion_relation} illustrates the power spectra of cyclotron waves from the fiducial simulation. The panels correspond to the two time intervals marked by gray bands in Figure \ref{fig:fiducial_isotropic_1d_perturbations_anisotropy_comparison}, representing the linear growth phases of the ion cyclotron ($0.6\,q^{-1}\lesssim t\lesssim0.8\,q^{-1}$; Figure \ref{fig:fiducial_dispersion_relation}(a)) and whistler ($1\,q^{-1}\lesssim t\lesssim1.2\,q^{-1}$; Figure \ref{fig:fiducial_dispersion_relation}(b)) instabilities. We output simulation data at a cadence of $\Delta t = 0.36\,\omega_{c,i0}^{-1}$ during these windows. We determine the dispersion relation $\omega(k_y)$ by computing the magnitude of the Fourier transform of the complex field $\delta B^z + i\delta B^x$ \citep{ley_stochastic_ion_acceleration_2019, tran_ic_icm_2023}. Because these modes propagate parallel to the mean field, we calculate the dispersion $\omega(k_y)$ after averaging the perturbations along the $x$ direction. To minimize spectral leakage, a Blackman window is applied in the temporal direction prior to the transform. The power spectrum is normalized by the number of elements in $\omega$ and $k_y$ to facilitate comparison between different time intervals. The Fourier transform of $\delta B^z + i\delta B^x$ allows us to separate left- and right-circularly polarized waves (LCP and RCP) based on the sign of $\omega$. For transverse perturbations proportional to $e^{i(k_yy-\omega t)}$ with $k_y>0$, LCP waves correspond to $\omega>0$ and RCP waves correspond to $\omega<0$ when viewed along the $+\hat{y}$ direction. In Figure \ref{fig:fiducial_dispersion_relation}(a), we observe the emergence of LCP modes associated with ion cyclotron waves satisfying $\omega>0$. The red curve in both panels of Figure \ref{fig:fiducial_dispersion_relation} represents the analytic dispersion relation for the ion cyclotron mode obtained by solving the 1D ($\boldsymbol{k}\parallel\boldsymbol{B}_0$) nonrelativistic dispersion equation for electromagnetic waves in a bi-Maxwellian electron-proton plasma \citep{davidson_ogden_ic_instability_1975,gary_madland_electron_anisotropies_1985,gary_ic_magnetosheath_theory_simulations_1993,guo_electron_heating_shocks_paper2_2018,tran_ic_icm_2023}.

Since the electrons are ultrarelativistic ($\Theta_e\gg1$), we leave a detailed linear theory comparison using a relativistic solver (e.g., ALPS; \citealt{verscharen_alps_2018}) for a future study. Nonetheless, we note several properties of the RCP whistler waves ($\omega<0$) in Figure \ref{fig:fiducial_dispersion_relation}(b). Their power is significantly weaker than that of the ion cyclotron waves, again indicating that whistler modes contribute subdominantly to the magnetic fluctuations, and it peaks at higher wavenumbers. The presence of comparable waves in the isotropic-ions simulation confirms that they are driven by electron anisotropy.

\begin{figure}
\centering
\includegraphics[width=\linewidth]{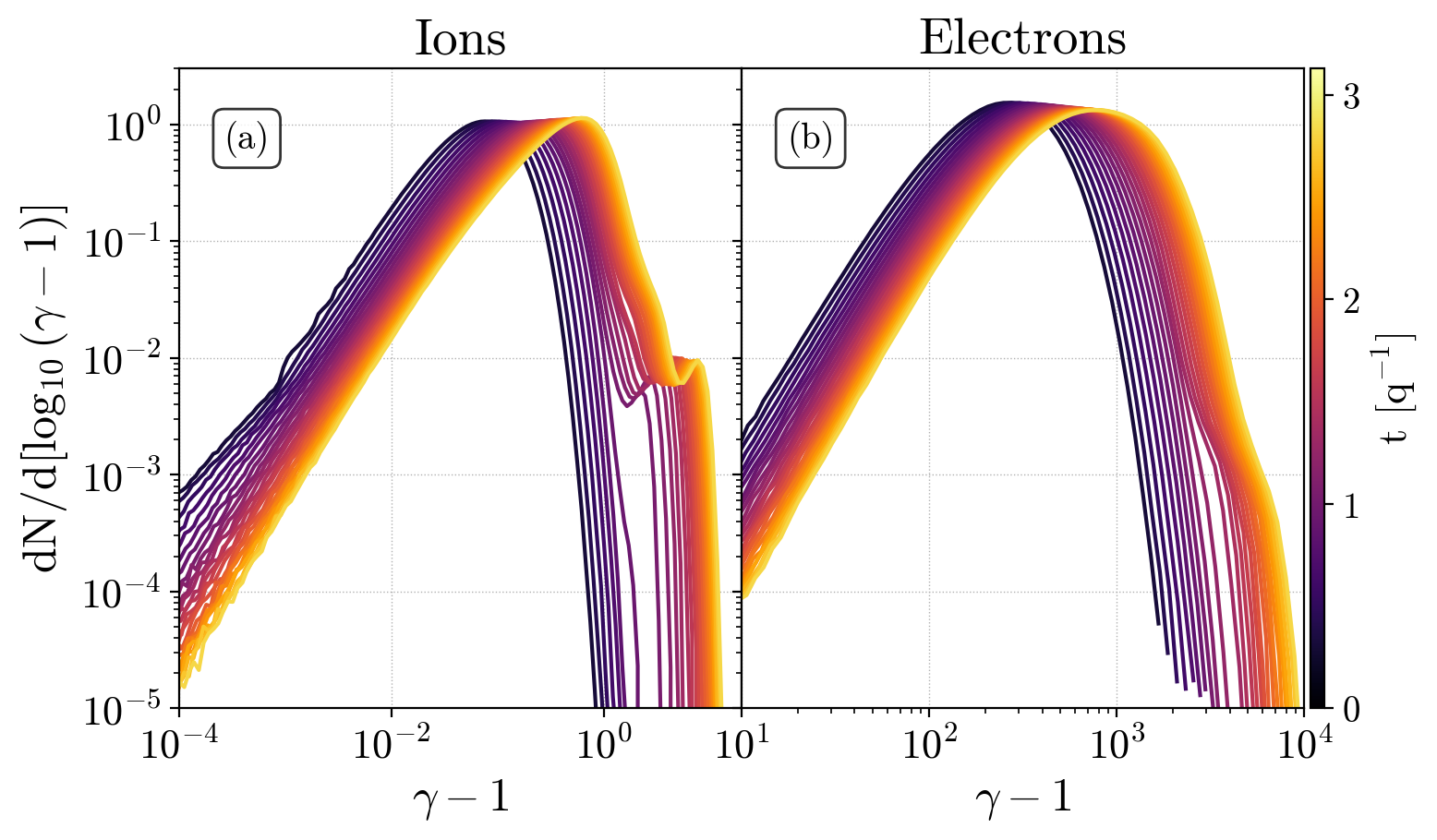}
\caption{Evolution of the ion (panel (a)) and electron (panel (b)) energy spectra for the fiducial simulation. The color gradient denotes the simulation time in units of $q^{-1}$, as shown by the colorbar. The initial spectrum for each species corresponds to an isotropic Maxwell-J\"{u}ttner distribution.}  
\label{fig:fiducial_energy_spectra}
\end{figure}

Figure \ref{fig:fiducial_energy_spectra} presents the time evolution of the energy spectra for ions (left panel) and electrons (right panel). Over the course of the simulation, the spectra shift to higher energies as the bulk plasma temperature increases. The ion spectrum develops a nonthermal tail ($1.4\lesssim\gamma_i-1\lesssim2$) and a high-energy nonthermal bump ($\gamma_i-1\simeq6$) as the ion cyclotron instability grows and saturates, with the quoted values corresponding to $t\simeq3\,q^{-1}$. Similarly, the electrons develop a nonthermal tail---albeit, a softer one---following the nonlinear onset of the whistler mode. While a detailed study of particle energization is the subject of future work, here we briefly comment on plausible heating and acceleration mechanisms, and on the relative roles of the various instabilities.

The development of parallel-propagating cyclotron waves can mediate stochastic particle acceleration through gyroresonant interactions, which can be modeled as diffusion in energy space \citep{melrose_resonant_scattering_wave_particle_solar_corona_1974, hamilton_stochastic_electron_acceleration_solar_flares_1992, dermer_stochastic_acceleration_black_hole_accretion_1996, petrosian_stochastic_acceleration_thermal_to_relativistic_2000}. The growth of fluctuating transverse magnetic fields generates associated transverse electric fields $\delta E_{\perp}\sim(\omega/kc)\,\delta B_{\perp}$ that perform work on the particles. Indeed, \cite{riquelme_stochastic_electron_acceleration_2017, ley_stochastic_ion_acceleration_2019} demonstrate using 2D shearing PIC simulations  that the work done by these fluctuating electric fields is the dominant heating mechanism for the nonthermal tail.


\begin{figure}
\centering
\includegraphics[width=\linewidth]{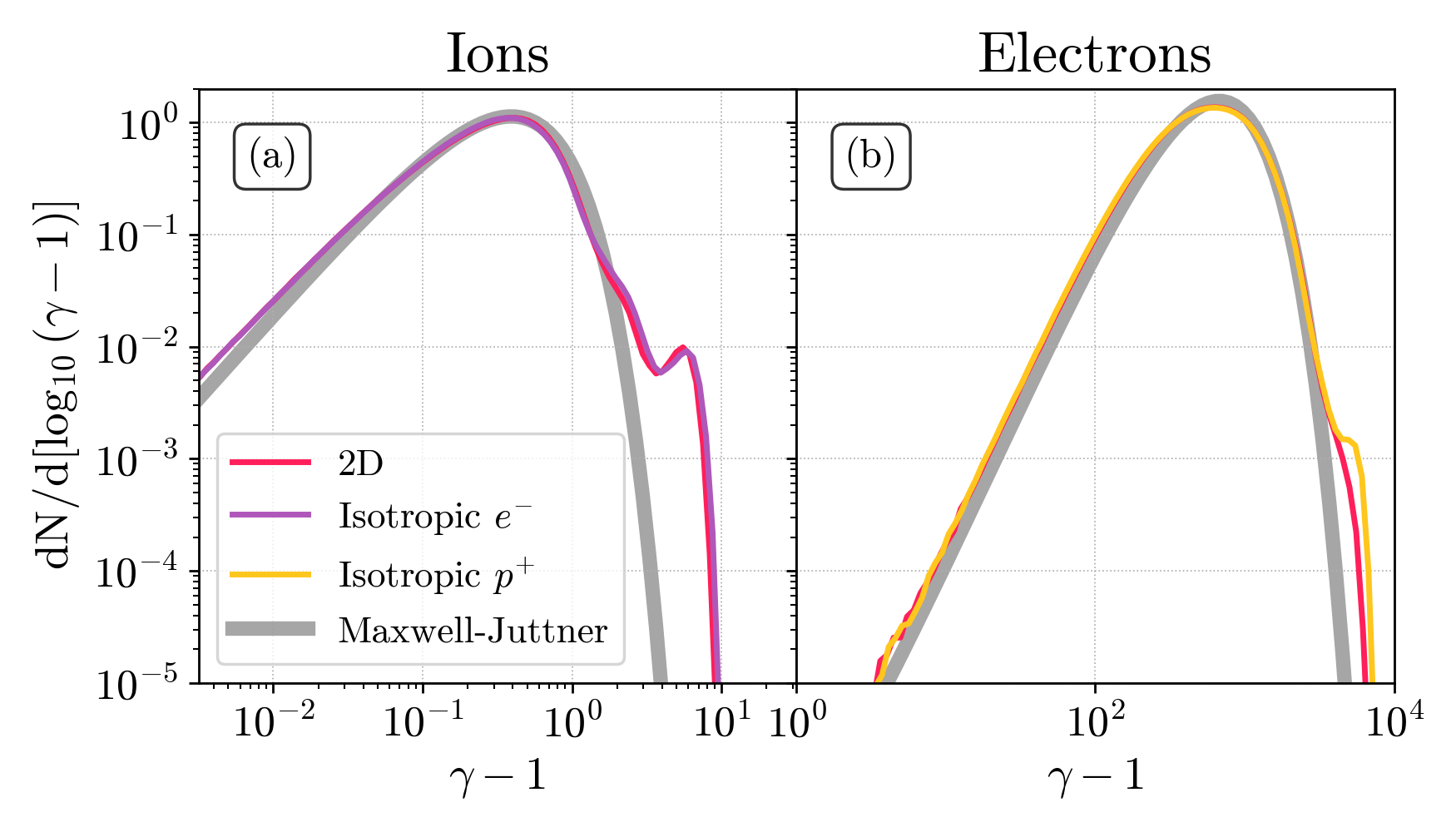}
\caption{Comparison of the ion (panel (a)) and electron (panel (b)) energy spectra for the fiducial simulation (red), the isotropic-electrons simulation (purple), and the isotropic-ions simulation (yellow). The spectra are plotted at $t=2\,q^{-1}$, when instabilities are saturated and well into the nonlinear phase. The gray curves in both panels represent the isotropic Maxwell-J\"{u}ttner distribution calculated using the effective temperature $\Theta_j = (2\Theta_{\perp,j} + \Theta_{\parallel,j})/3$ at this time.} 
\label{fig:fiducial_energy_spectra_comparison_with_isotropic}
\end{figure}

Figure \ref{fig:fiducial_energy_spectra_comparison_with_isotropic} compares the particle energy spectra from the fiducial simulation (red) with those from equivalent simulations employing isotropic electrons (purple) and isotropic ions (yellow). This comparison enables us to isolate the role of the mirror instability in particle energization, as mirror modes are suppressed in the isotropic cases. The ion spectrum from the isotropic-electron simulation exhibits excellent agreement with the fiducial case, suggesting that ion cyclotron modes are the primary driver of the nonthermal tail. Because mirror modes are nonpropagating ($\omega=0$), they generate negligible electric fields and thus contribute little to ion energization. This finding is consistent with the broader results of this subsection, which point to a limited role of mirror modes in ion dynamics. In contrast, mirror modes appear to slightly suppress electron energization, as indicated by the deviation between the red and yellow curves in Figure \ref{fig:fiducial_energy_spectra_comparison_with_isotropic}(b).

Macroscopic velocity gradients, such as compression and shear, can produce irreversible bulk heating by generating a pressure anisotropy that is then capped by collisions or effective pitch-angle scattering. This dissipation is referred to variously in the literature as ``frictional heating due to nonuniform velocities'' \citep{kulsrud_mhd_plasma_1983}, ``volumetric viscous heating'' \citep{hollweg_viscosity_magnetized_plasma_1985}, or ``parallel viscous heating'' \citep{kunz_icm_heating_mechanism_2011}. In the specific context of repeated compression–expansion cycles, the same mechanism is often termed ``magnetic pumping'' \citep{berger_magnetic_pumping_1958, lichko_magnetic_pumping_2017, ley_magnetic_pumping_2023, tran_ic_icm_2023}. Resonant wave-particle interactions provide pitch-angle scattering, which introduces a phase lag between the pressure anisotropy and the compression-driven evolution of the magnetic field. When the scattering rate is comparable to the compression rate, this drives net particle energization. Though we have shown that mirror modes do not directly energize ions via stochastic acceleration, both mirror modes' and ion cyclotron waves' pitch-angle scattering may drive magnetic pumping of ions. The net energy gain from pumping is not visible in Figures~\ref{fig:fiducial_energy_spectra}--\ref{fig:fiducial_energy_spectra_comparison_with_isotropic}, but it would be realized upon completing a compression-expansion cycle, where the mean magnetic field goes back to its initial value.

\section{Dependence of Wave Properties on Plasma Conditions}
\label{sec:results_parameter_analysis}

We now examine the trends in the evolution of dominant plasma instabilities and particle acceleration as a function of the initial simulation parameters. We systematically investigate the dependence on ion plasma beta $\beta_{i0}$ (Section \ref{sec:results_ion_plasma_beta}), ion temperature $\Theta_{i0}$ (Section \ref{sec:results_ion_temperature}), electron-to-ion temperature ratio $T_{e0}/T_{i0}$ (Section \ref{sec:results_electron_to_ion_temperature_ratio}), and compression rate $\omega_{c,i0}/q$ (Section \ref{sec:results_compression_rate}).

\subsection{Dependence on Initial Ion Plasma Beta $\beta_{i0}$}
\label{sec:results_ion_plasma_beta}

\begin{figure*}
\centering
\includegraphics[width=\textwidth]{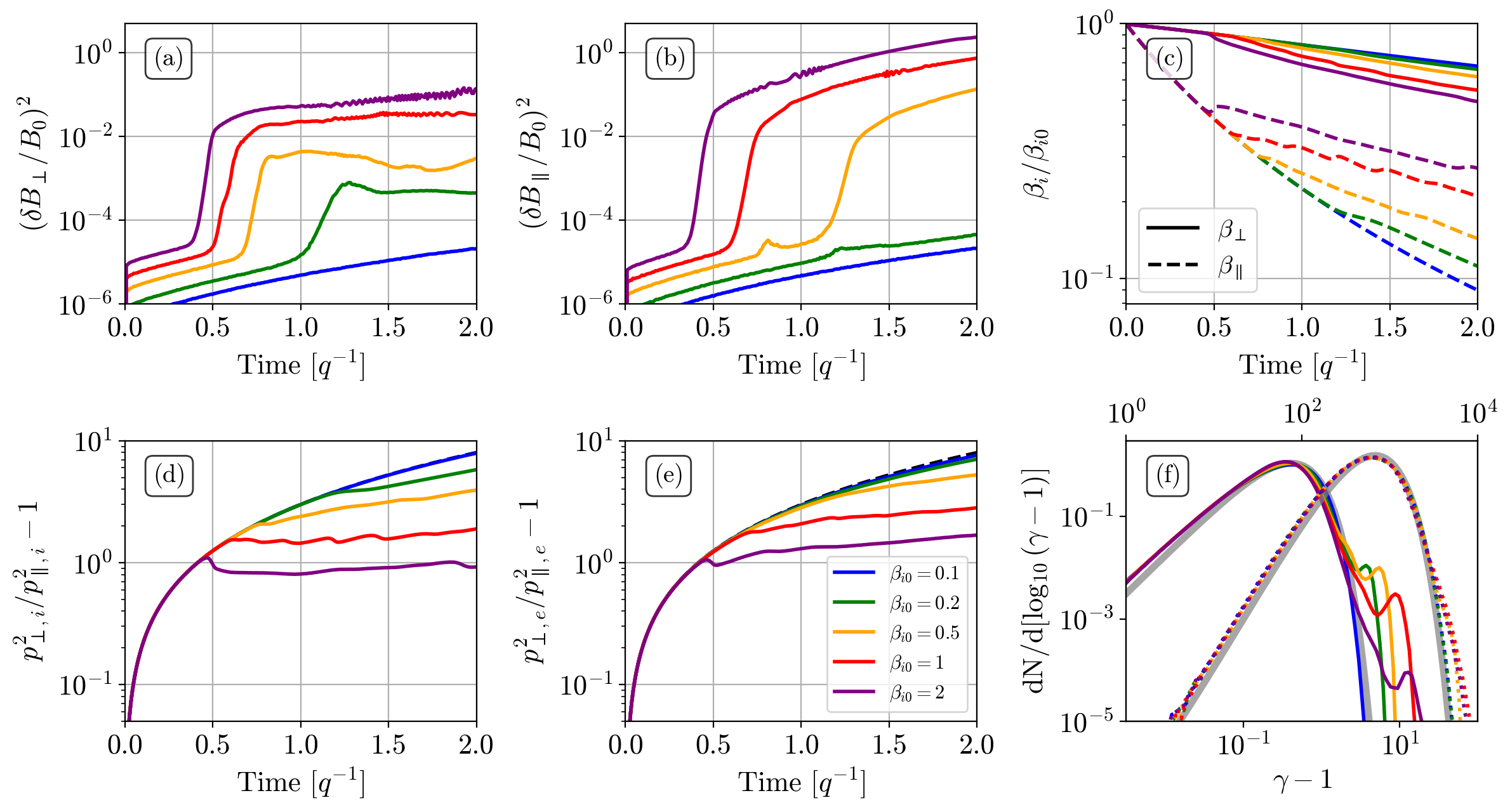}
\caption{Time evolution of plasma instabilities in a compressing box as a function of $\beta_{i0}$ (fixing $T_{e0}/T_{i0}=1$ and $\Theta_{i0}=0.05$).  We consider five values for $\beta_{i0}$: 0.1 (blue), 0.2 (green), 0.5 (orange), 1 (red), and 2 (purple). Panels (a) and (b) display the transverse and longitudinal magnetic field perturbations normalized by $B_0$, respectively. Panel (c) shows the evolution of perpendicular (solid) and parallel (dashed) ion plasma beta normalized by $\beta_{i0}$. Panels (d) and (e) plot the temporal evolution of ion and electron anisotropy, respectively. Panel (f) presents the ion (solid) and electron (dotted) energy spectra at $t=2\,q^{-1}$. The bottom $x$-axis in (f) corresponds to ions, while the top $x$-axis applies to electrons. Dashed black lines in (d)--(e) indicate the expected adiabatic evolution $(1 + qt)^2-1$. Gray curves in panel (f) represent the isotropic Maxwell-J\"{u}ttner distributions for the respective species.}
\label{fig:ion_beta_study_times_series_and_spectra}
\end{figure*}

For fixed $\Theta_{i0}$ and $T_{e0}/T_{i0}$, the initial ion plasma beta $\beta_{i0}$ determines the mean field strength $B_0$. Therefore, varying $\beta_{i0}$ allows us to assess the dependence of our results on the background magnetic field, or equivalently on the initial ion Alfv\'{e}n velocity $v_{A,i0}/c$ at fixed $\Theta_{i0}$.

In Figure \ref{fig:ion_beta_study_times_series_and_spectra}, we analyze the development of plasma instabilities across a range of $\beta_{i0}$ values centered on the fiducial case: 0.1 (blue), 0.2 (green), 0.5 (orange), 1 (red), and 2 (purple). We observe that the ion cyclotron instability onsets earlier as $\beta_{i0}$ increases, as evidenced by the evolution of $(\delta B_{\perp}/B)^2$ and $A_i$ in Figures \ref{fig:ion_beta_study_times_series_and_spectra}(a,d). This is consistent with a lower ion cyclotron anisotropy threshold at higher $\beta_{\parallel,i}$. A higher $\beta_{i0}$ implies a weaker restoring magnetic force, so a lower anisotropy is sufficient to destabilize the plasma. The mirror instability follows a similar trend, as shown by the evolution of $(\delta B_{\parallel}/B)^2$ in Figure \ref{fig:ion_beta_study_times_series_and_spectra}(b). This aligns with the expectation that the mirror anisotropy threshold scales as $\propto1/\beta_{\perp,i}$. For $\beta_{i0}=0.1,0.2$, we do not observe the growth of the mirror mode. For $\beta_{i0}=1$ and $2$, the mirror instability develops concurrently with, or immediately following, the ion cyclotron instability. Mirror fluctuations reach larger amplitudes than ion cyclotron fluctuations by the end of the simulations. As a result, the late-time ion anisotropy is modified by the formation of mirror structures. This is supported by comparison with corresponding 1D simulations, in which mirror modes are absent.

The threshold for the relativistic whistler instability is parameterized as $A_{\text{T},e}=S_e/\beta_{\perp,e}^{\alpha_e}$ \citep[Appendix B.3]{galishnikova_anisotropic_images_2023}, where $\alpha_e\lesssim1$ and $S_e<1$ with a weak dependence on $\Theta_e$. This implies a higher anisotropy threshold for simulations with lower $\beta_{i0}$ (given $\beta_{e0}=\beta_{i0}$), consistent with the $A_e$ evolution shown in Figure \ref{fig:ion_beta_study_times_series_and_spectra}(e). For the $\beta_{i0}=1$ and $2$ cases, which exhibit the earliest onset of the mirror instability, the late-time electron anisotropy (by $t=2\,q^{-1}$) is governed by the secular growth of the mirror mode, as in the fiducial case. We establish this by comparing with corresponding isotropic-ions simulations, in which mirror modes are absent. Therefore, for $\beta_{i0}\geq0.5$, mirror modes influence the late-time electron anisotropy. 

The temporal evolution of $\beta_{\parallel,i}/\beta_{i0}$ and $\beta_{\perp,i}/\beta_{i0}$, is shown in Figure \ref{fig:ion_beta_study_times_series_and_spectra}(c). The secular growth of the mean field ($\propto(1+qt)^2$) drives a gradual reduction in $\beta_i$ (more pronounced for $\beta_{\parallel,i}$), highlighting that plasma conditions evolve in time and that the onset of microinstabilities is governed by their \textit{instantaneous} values. For $\beta_{i0}=0.1$, the instability thresholds are sufficiently high that no modes are excited over the full duration of our simulations, and both ions and electrons evolve adiabatically (see the blue curve in Figures \ref{fig:ion_beta_study_times_series_and_spectra}(d,e)).

As discussed for the fiducial case, particles can undergo stochastic acceleration via gyroresonant interactions with their respective cyclotron waves, producing nonthermal tails. Figure \ref{fig:ion_beta_study_times_series_and_spectra}(f) tracks this energization across varying $\beta_{i0}$ at $t=2\,q^{-1}$. We observe nonthermal components in both species for $\beta_{i0}=0.5,1,2$, where cyclotron instabilities are observed to develop. At higher $\beta_{i0}$, the mirror instability impedes efficient ion scattering by ion cyclotron waves, resulting in softer ion spectra as $\beta_{i0}$ increases. The electron spectra exhibit a similar nonthermal tail in all cases where $A_e$ departs from the adiabatic expectation.

\subsection{Dependence on Initial Ion Temperature $\Theta_{i0}$}
\label{sec:results_ion_temperature}

\begin{figure*}
\centering
\includegraphics[width=\textwidth]{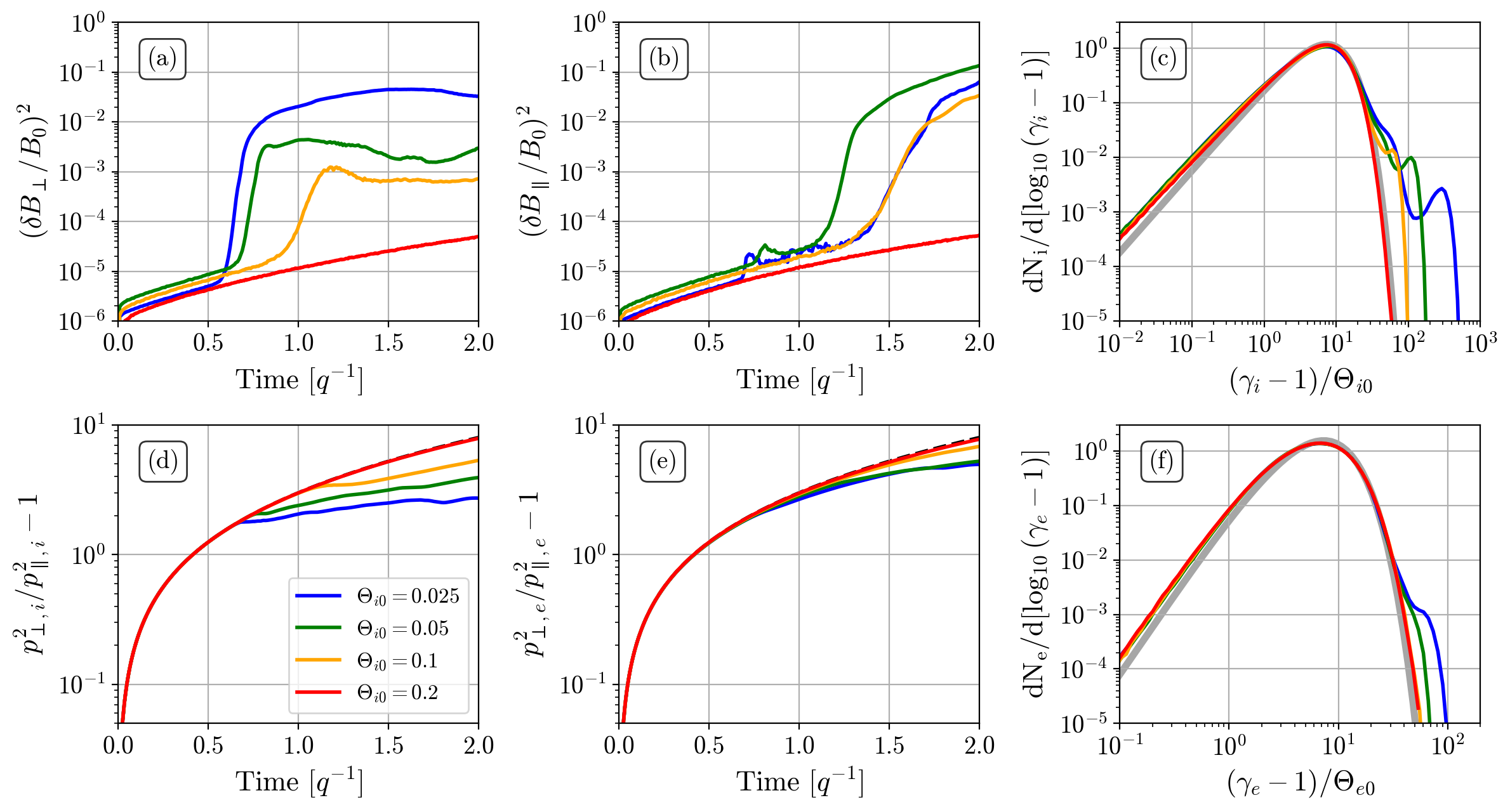}
\caption{Time evolution of plasma instabilities in a compressing box as a function of $\Theta_{i0}$ (fixing $T_{e0}/T_{i0}=1$ and $\beta_{i0}=0.5$).  We consider four values for $\Theta_{i0}$: 0.025 (blue), 0.05 (green), 0.1 (orange), and 0.2 (red). Panels (a) and (b) display the transverse and longitudinal magnetic field perturbations normalized by $B_0$, respectively. Panel (c) shows the ion energy spectra at $t=2\,q^{-1}$. Panels (d) and (e) plot the temporal evolution of ion and electron anisotropy, respectively. Panel (f) presents electron energy spectra at $t=2\,q^{-1}$. Dashed black lines in (d)--(e) indicate the expected adiabatic evolution $(1 + qt)^2-1$. Gray curves in panels (c,f) represent the isotropic Maxwell-J\"{u}ttner distribution for $\Theta_{i0}=0.2$.}
\label{fig:thetai_study_beta0.5_times_series_and_spectra}
\end{figure*}


In this section we examine how the growth and saturation of anisotropy-driven instabilities depend on the initial ion temperature $\Theta_{i0}$. Because $\Theta_{e0} = \Theta_{i0}\cdot(m_i/m_e)$ for the fiducial case ($T_{e0}=T_{i0}$), varying $\Theta_{i0}$ simultaneously varies the thermal content---and hence the degree of relativistic effects---of both species. This scan therefore also probes how instabilities driven by electron anisotropy depend on $\Theta_{e0}$. For fixed $\beta_{i0}$ and $T_{e0}/T_{i0}$, the scan can equivalently be expressed in terms of the initial ion Alfv\'{e}n velocity, since $\Theta_{i0}=\beta_{i0}/2\cdot(v_{A,i0}/c)^2$ (assuming nonrelativistic ions). \cite{sironi_electron_heating_2015} report little dependence on $\Theta_{i0}$ for the parameters explored there; however, their study lies in a nonrelativistic regime (using their fiducial $v_{A,i0}/c=0.05$, $\beta_{i0}=20$, $m_i/m_e=16$, and $T_{e0}/T_{i0}=10^{-2}$, implying $\Theta_{i0}=0.025$ and $\Theta_{e0}=0.004$), whereas here we focus on the  transrelativistic regime, most relevant for MAD flows.

Linear analyses of parallel-propagating, RCP whistler waves driven by relativistically hot, anisotropic electrons show that the instability weakens as the electrons are made more relativistic. In particular, \cite{gladd_relativistic_whistler_1983, xiao_relativistic_whistler_growth_1998} find that the whistler growth rate decreases with increasing $\Theta_{\perp,e}$. \cite{xiao_relativistic_whistler_growth_1998} attribute this reduction to relativistic modifications of the cyclotron resonance, which diminishes both the effective pitch-angle anisotropy of the resonant electrons and the fraction of resonant electrons. \cite{gladd_relativistic_whistler_1983} compute marginal stability curves and find that the electron anisotropy threshold increases with $\Theta_{\perp,e}$. It is reasonable to anticipate an analogous trend for relativistic ions interacting with LCP ion cyclotron waves (given the cyclotron nature of both modes): as $\Theta_{i0}$ increases, the linear growth rate decreases and the anisotropy threshold to trigger ion cyclotron waves increases. 


Figure \ref{fig:thetai_study_beta0.5_times_series_and_spectra} shows the $\Theta_{i0}$ scan for $T_{e0}/T_{i0}=1$ and $\beta_{i0}=0.5$. From the time evolution of the transverse magnetic perturbations (panel (a)) and the ion anisotropy $A_i$ (panel (d)), we find that the onset of the ion cyclotron mode shifts to later times as $\Theta_{i0}$ increases. This trend reflects the increase in the ion cyclotron anisotropy threshold with $\Theta_{\perp,i}$. In the $\Theta_{i0}=0.2$ run, the ions evolve adiabatically over the entire duration of the run, suggesting that the anisotropy threshold for ion cyclotron or mirror instability is never attained. Electrons exhibit a similar dependence on the initial temperature: panel (e) shows that the electron anisotropy at marginal stability increases with $\Theta_{i0}$. However, the simultaneous presence of mirror activity makes it difficult to unambiguously attribute the electron anisotropy at late times to the whistler anisotropy threshold. Otherwise identical 2D isotropic-ions simulations isolate the whistler contribution because mirror modes are absent in these runs. They show that (i) the whistler onset shifts to later times as $\Theta_{e0}$ increases and (ii) the whistler anisotropy threshold increases with $\Theta_{e0}$, consistent with expectations from relativistic linear theory. Comparing $A_e$ in these isotropic-ions runs with the corresponding 2D runs in which both species develop anisotropy shows that the influence of mirror modes on the electron anisotropy reduces as $\Theta_{i0}$ increases. Finally, the ion and electron energy spectra (panels (c) and (f)) suggest that nonthermal acceleration is more pronounced at lower $\Theta_{i0}$.

\subsection{Dependence on Initial Electron-to-Ion Temperature ratio $T_{e0}/T_{i0}$}
\label{sec:results_electron_to_ion_temperature_ratio}

\begin{figure*}
\centering
\includegraphics[width=0.9\textwidth]{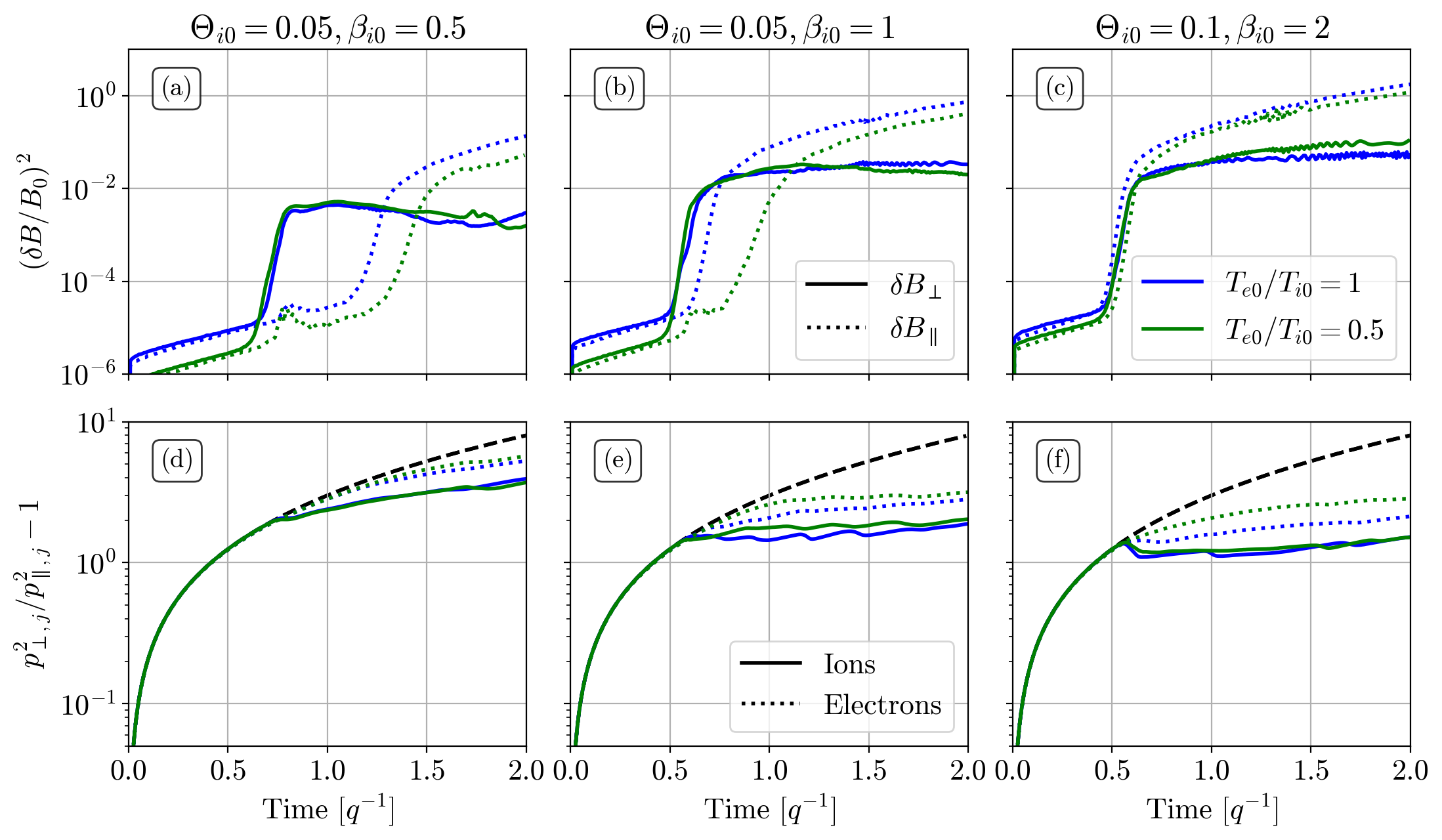}
\caption{Dependence on the initial electron-to-ion temperature ratio $T_{e0}/T_{i0}$. We compare two ratios, $1$ (blue) and $0.5$ (green), for simulations initialized with: $\Theta_{i0}=0.05,\,\beta_{i0}=0.5$ (left column; our fiducial case); $\Theta_{i0}=0.05,\,\beta_{i0}=1$ (middle column); and $\Theta_{i0}=0.1,\,\beta_{i0}=2$ (right column). Panels (a)--(c) display the transverse (solid) and longitudinal (dotted) magnetic field perturbations. Panels (d)--(f) show the ion (solid) and electron (dotted) anisotropies; the dashed black line indicates the expected adiabatic evolution.}
\label{fig:temperature_ratio_study_times_series}
\end{figure*}

\begin{figure*}
\centering
\includegraphics[width=0.9\textwidth]{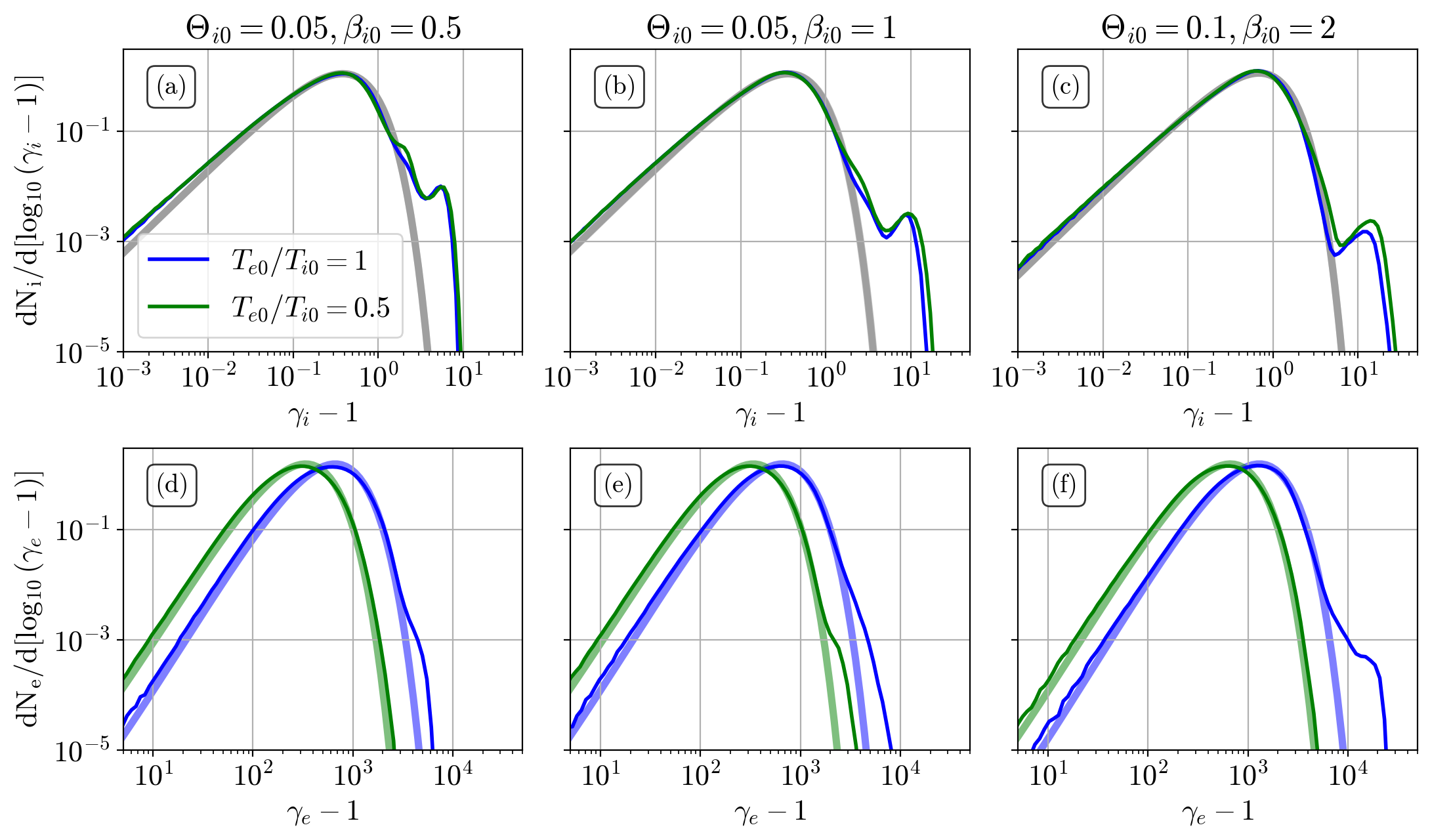}
\caption{Dependence on the initial electron-to-ion temperature ratio $T_{e0}/T_{i0}$. The simulations correspond to those shown in Figure \ref{fig:temperature_ratio_study_times_series}. Panels (a)--(c) show the ion energy spectra and panels (d)--(f) show the electron energy spectra. In panels (a)--(c), the gray curves indicate the isotropic Maxwell-J\"{u}ttner ion distribution. In panels (d)--(f), the semi-transparent blue and green curves indicate the isotropic Maxwell-J\"{u}ttner electron distributions for $T_{e0}/T_{i0}=1$ and $0.5$, respectively.}
\label{fig:temperature_ratio_study_spectra}
\end{figure*}

In this section, we investigate the influence of  $T_{e0}/T_{i0}$ on the onset of plasma instabilities and the regulation of particle anisotropy. Due to the computational cost of simulating a two-temperature plasma where $T_{i}\gg T_{e}$ while maintaining a realistic mass ratio $m_i/m_e$, we limit our study to $T_{e0}/T_{i0} \geq 0.5$. A larger temperature difference between the two species increases the rate of numerical heating, necessitating higher \texttt{ppc}.


Figures \ref{fig:temperature_ratio_study_times_series} and \ref{fig:temperature_ratio_study_spectra} illustrate the effect of varying the initial electron temperature across three simulations: $\Theta_{i0}=0.05,\,\beta_{i0}=0.5$ (the fiducial case; left column); $\Theta_{i0}=0.05,\,\beta_{i0}=1$ (middle column); and $\Theta_{i0}=0.1,\,\beta_{i0}=2$ (right column). In Figures \ref{fig:temperature_ratio_study_times_series}(a)--(c), the solid lines show the time evolution of $(\delta B_{\perp}/B_0)^2$ for $T_{e0}/T_{i0}=1$ (blue) and $0.5$ (green). For the simulations in Figures \ref{fig:temperature_ratio_study_times_series}(a,b), the initial electron temperature has little effect on the transverse perturbations until $t\simeq1.5\,q^{-1}$. During this phase, $(\delta B_{\perp}/B_0)^2$ is set by the ion cyclotron mode, after which the mirror mode dominates ($t\gtrsim1.5\,q^{-1}$). For the $\beta_{i0}=2$ case (Figure \ref{fig:temperature_ratio_study_times_series}(c)), the transverse perturbations are dominated by mirror fluctuations from $t\gtrsim0.6\,q^{-1}$. Here, the onset times of the mirror instability are comparable between the two $T_{e0}/T_{i0}$ cases, and the mode amplitudes overlap closely during the secular growth phase. The longitudinal perturbations (dashed lines in Figures \ref{fig:temperature_ratio_study_times_series}(a)--(c)) indicate that the onset of the mirror instability is delayed for $T_{e0}/T_{i0}=0.5$. This is consistent with the theoretical expectation that lowering the electron temperature, at fixed $A_{\rm{T},e}$, raises the mirror instability threshold and suppresses its growth rate when $T_{\perp,e}>T_{\parallel,e}$ \citep{pokhotelov_electron_anisotropy_mirror_2000}.

Figure \ref{fig:temperature_ratio_study_times_series}(d) shows that the ion anisotropy follows the same trajectory regardless of $T_{e0}/T_{i0}$ for the fiducial values of $\beta_{i0}$ and $\Theta_{i0}$, suggesting that the evolution of the ion cyclotron instability---and its regulation of $A_i$---is largely unaffected by the initial electron temperature. At higher $\beta_{i0}$, where the mirror instability regulates the late-time ion anisotropy, $A_i$ remains nearly identical across the probed ratios (Figures \ref{fig:temperature_ratio_study_times_series}(e,f)). This suggests that, for the parameters explored here, electron contributions to the ion mirror threshold are negligible when $\beta_{i0}\gtrsim1$. We find that at late times ($t=2\,q^{-1}$), the electron anisotropy in all cases is regulated by mirror modes, saturating at a higher value for the simulations with initially colder electrons ($T_{e0}/T_{i0}=0.5$). We confirm the role of mirror fluctuations in setting $A_e$ by comparing our 2D results for the $T_{e0}/T_{i0}=0.5$ simulations against equivalent 1D runs where mirror instabilities cannot develop. The electron anisotropy at $t=2\,q^{-1}$ in the 1D runs remains persistently higher than in the corresponding 2D cases.

In Figures \ref{fig:temperature_ratio_study_spectra}(a)--(c) and (d)--(f), we plot the ion and electron energy spectra respectively at $t=2\,q^{-1}$. The ion spectra show agreement between the simulations, indicating that ion energization is insensitive to the initial electron thermal content. The electron spectra for the reduced temperature ratio cases are shifted to lower energies due to their lower initial thermal energy. We observe that a nonthermal tail develops more prominently in the $T_{e0}=T_{i0}$ simulations. This is likely because the generated whistler waves are weaker at lower electron temperatures. 1D isotropic-ions simulations confirm the dependence of the whistler fluctuation amplitude on $T_{e0}/T_{i0}$.

\begin{figure*}
\centering
\includegraphics[width=\textwidth]{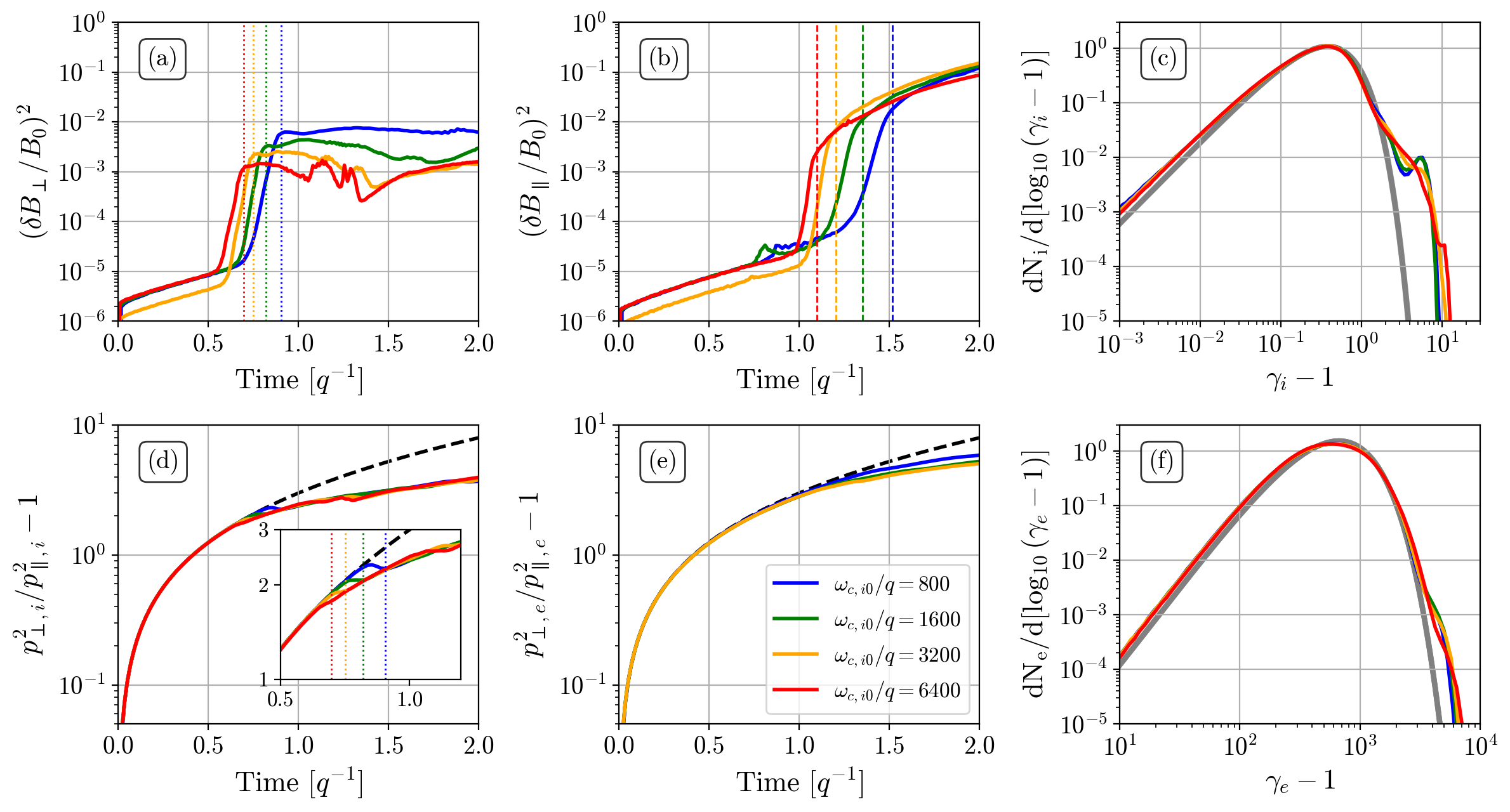}
\caption{Time evolution of plasma instabilities in a compressing box as a function of compression rate. We consider four values of $\omega_{c,i0}/q$: 800 (blue), 1600 (green), 3200 (orange), and 6400 (red). Panels (a) and (b) display the transverse
and longitudinal magnetic field perturbations normalized by $B_0$, respectively. Panel (c) shows the ion spectra at $t=2\,q^{-1}$. Panels (d) and (e) plot the temporal evolution of ion and electron anisotropy, respectively. Panel (f) presents the electron energy spectra at $t=2\,q^{-1}$. Vertical dotted lines in panel (a) mark the end of the exponential phase of the ion cyclotron instability, coincident with the saturation of ion anisotropy. As shown in the inset of panel (d) for the interval $t=0.5\text{--}1\,q^{-1}$, these same dotted lines mark when significant pitch-angle scattering causes $A_i$ to deviate from the adiabatic prediction and saturate at the anisotropy threshold. The vertical dashed lines in panel (b) mark the end of exponential mirror growth. Dashed black lines in panels (d,e) indicate the expected adiabatic evolution $(1 + qt)^2-1$. We do not plot $A_e$ for the $\omega_{c,i0}/q=6400$ case because the \texttt{ppc} is insufficient to reliably capture the electron dynamics (see Footnote \ref{footnote:6400_compression_rate_case}). Gray curves in panels (c,f) represent the isotropic Maxwell-J\"{u}ttner distribution.}
\label{fig:compression_rate_study_times_series_and_spectra}
\end{figure*}

\subsection{Dependence on Compression Rate $\omega_{c,i0}/q$}
\label{sec:results_compression_rate}

In all simulations presented thus far, we have held the compression rate fixed at $\omega_{c,i0}/q=1600$. We now investigate the impact of varying this parameter, specifically examining how it influences the onset and saturation amplitude of the instabilities, the ion and electron anisotropies at marginal stability, and the energy spectra during the secular phase.

Figure \ref{fig:compression_rate_study_times_series_and_spectra}(a) plots the time evolution of the transverse magnetic field perturbations for different compression rates. The growth rate of the ion cyclotron instability is set by the background compression rate, which governs how rapidly the ion anisotropy is driven by magnetic field amplification and the conservation of adiabatic invariants. We observe that the ion cyclotron instability onsets earlier for larger $\omega_{c,i0}/q$ (i.e., slower compression), when measured in units of $q^{-1}$. This is because, for faster compression, the ions are pushed further above marginal stability (see the inset in Figure \ref{fig:compression_rate_study_times_series_and_spectra}(d)), before the growing ion cyclotron fluctuations attain sufficient amplitude to pitch-angle scatter the ions and relax $A_i$ back toward its threshold value (see \citealp{hellinger_magnetosheath_compression_2005, riquelme_ic_shearing_pic_2015,sironi_electron_heating_2015} for related discussions in compression and shear setups). The dotted vertical lines in Figure \ref{fig:compression_rate_study_times_series_and_spectra}(a) and in the inset in Figure \ref{fig:compression_rate_study_times_series_and_spectra}(d) mark the end of the linear phase of the ion cyclotron instability. The difference in the instability onset time between progressively slower compressions decreases, indicating that the onset converges to a finite limit in the regime $\omega_{c,i0}/q \gg 1$. We confirm this with 1D simulations, which allow us to probe slower compression rates (up to $\omega_{c,i0}/q=12800$).

The evolution of the mirror instability is similarly sensitive to the compression rate. Figure \ref{fig:compression_rate_study_times_series_and_spectra}(b) displays the longitudinal field perturbations, demonstrating that the mirror instability activates earlier as $\omega_{c,i0}/q$ increases. For the fiducial parameters ($\beta_{i0}=0.5,\,\Theta_{i0}=0.05,\,T_{e0}/T_{i0}=1$), we find that the saturation amplitude of the mirror mode appears insensitive to the compression rate\footnote{\label{footnote:6400_compression_rate_case}The saturated field strength $(\delta B_{\parallel}/B_0)^2$ is slightly lower for the $\omega_{c,i0}/q=6400$ case. We attribute this to insufficient particle statistics, which limits our ability to accurately resolve the electron dynamics. Because the mirror instability is sourced by the anisotropy of both ions and electrons, errors in the electron contribution reduce the effective drive, thereby lowering the saturation amplitude. Although the level of noise is not sufficient to properly quantify the degree at which electron anisotropy departs from the adiabatic expectation, it remains true that electrons are highly anisotropic; as such, their contribution to the mirror mode is at zeroth order properly captured.}. \cite{Kunz_2014, riquelme_ic_shearing_pic_2015} report an analogous trend in shearing-sheet simulations of high-$\beta$ ($20\lesssim\beta\lesssim200$) plasmas: the saturated amplitude of the mirror-driven magnetic fluctuations is largely independent of the imposed shear rate.

We find that the whistler instability exhibits the same compression rate dependence as the ion cyclotron instability: slower compression leads to an earlier onset in units of $q^{-1}$. Consequently, the electron anisotropy departs from adiabatic evolution earlier in the more slowly compressed boxes, as shown in Figure \ref{fig:compression_rate_study_times_series_and_spectra}(e). We verify this interpretation using matched 1D simulations, in which oblique mirror modes are suppressed and parallel-propagating whistler waves regulate the electron anisotropy. We also find that the saturated electron anisotropy appears largely insensitive to the compression rate.

Figures \ref{fig:compression_rate_study_times_series_and_spectra}(c,f) present the ion and electron energy spectra at $t=2\,q^{-1}$. Both species exhibit nonthermal tails across all compression rates. For the ions, slower compression produces a nonthermal tail resembling a power law over $1.5\lesssim\gamma_i-1\lesssim7$, whereas faster compression generates a distinct nonthermal bump near $\gamma_i-1\simeq6$. The overall ion spectral shape, however, remains consistent with the fiducial simulation. We notice an identical trend in equivalent 1D simulations: faster compression develops a pronounced bump at $\gamma_i-1\simeq6$, while slower compression yields a power law tail. The electron spectra in Figure \ref{fig:compression_rate_study_times_series_and_spectra}(f) exhibit little sensitivity to the compression rate. In analogous 1D simulations, however, faster compression produces a more prominent nonthermal bump in the electron spectra. We interpret the weak compression rate dependence of the electron spectra in 2D as a consequence of mirror fluctuations regulating whistler activity and limiting the stochastic acceleration responsible for the electron tail.
\section{Summary}
\label{sec:summary}

\begin{figure*}
\centering
\includegraphics[width=\textwidth]{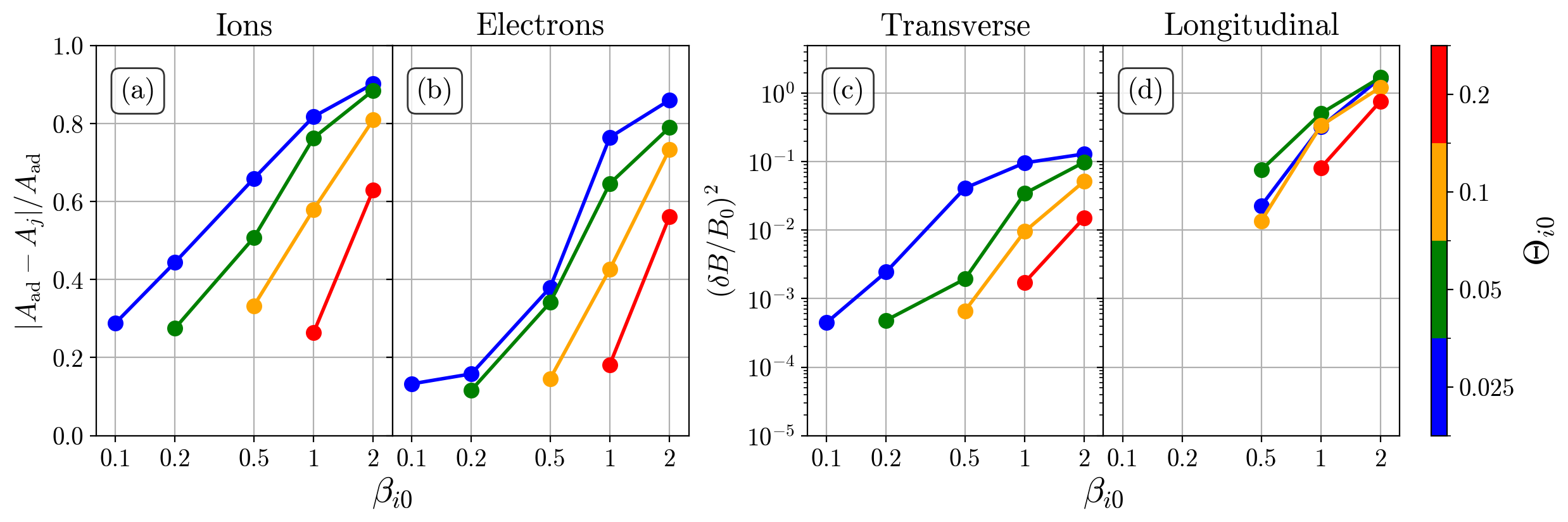}
\caption{Late-time diagnostics for 2D simulations with $T_{e0}/T_{i0}=1$ and $\omega_{c,i0}/q=1600$. Panels (a) and (b) show the fractional deviation of ion and electron anisotropy, $\Delta_j\equiv (A_{\text{ad}} - A_j)/A_{\text{ad}}$, from the adiabatic expectation at $t=2\,q^{-1}$; $\Delta_j=0$ indicates adiabatic evolution. Recall that particle anisotropy is defined as $A_j\equiv p_{\perp,j}^2/p_{\parallel,j}^2-1$, where $p_{\perp,j}^2$ and $p_{\parallel,j}^2$ are box-averaged momentum-squared components perpendicular and parallel to $\boldsymbol{B}_0$. Panels (c) and (d) show the transverse and longitudinal magnetic field perturbations, averaged over $t=1.5$--$2\,q^{-1}$. Colors denote different $\Theta_{i0}$ values. We omit plotting $\Delta_j$ for simulations where no instabilities develop, $\delta B_{\perp}$ for cases lacking cyclotron instabilities, and $\delta B_{\parallel}$ for cases lacking mirror growth.}
\label{fig:simulation_suite_late_time_anisotropy_perturbations}
\end{figure*}

In this work we study the development of kinetic instabilities in a low-$\beta$, transrelativistic plasma ($\Theta_i\lesssim0.1$ and $\Theta_e\gg1$), motivated by the plasma conditions expected in the inner few gravitational radii of magnetically arrested disks (MADs). We perform fully kinetic 2D (and selected 1D) simulations of an electron–ion plasma undergoing compression perpendicular to a mean magnetic field. The plasma parameters adopted in our simulations are motivated by the local conditions in regions of the flow that dominate the emission in synthetic \sgra images from weakly collisional MAD models \citep{dhruv_egrmhd_variability_2025}. Similarly, the predominance of $P_{\perp}>P_{\parallel}$ in the weakly collisional models motivates our use of a compressing-box setup, which drives anisotropy of the same sign. The adopted parameters allow simulations with the realistic mass ratio, $m_i/m_e=1836$, which we use in all runs presented here. This is feasible because relativistic electrons reduce the effective mass ratio to $m_i/\langle\gamma_e\rangle m_e\lesssim 10$. Due to computational constraints, we employ a compression timescale of $ \omega_{c,i0}/q \sim 10^{3}\text{--}10^{4} $, much smaller than in realistic accretion flows. The fiducial case has parameters $\beta_{i0}=0.5$, $\Theta_{i0}=0.05$ and $T_{e0}/T_{i0}=1$, and we explore the surrounding parameter space to quantify how the instability evolution depends on the flow conditions. Our main results are as follows:\begin{enumerate}
    \item In the fiducial simulation, the ion cyclotron instability develops before the mirror instability, and the evolution of ion anisotropy is largely regulated through pitch-angle scattering off ion cyclotron waves even as mirror fluctuations grow secularly. This suggests that future global studies that evolve $\Delta P$ should account for the ion cyclotron threshold when applying anisotropy limiters. The electron anisotropy drives whistler waves, leading to departure from adiabatic evolution. However, we find that the late-time electron anisotropy is affected by mirror modes. The mirror instability is nonresonant and driven by both ion and electron anisotropy \citep{pokhotelov_electron_anisotropy_mirror_2000,hellinger_mirror_multispecies_2007, remya_electron_anisotropy_ic_mirror_2013}. We observe mirror growth only when both species develop anisotropy; in otherwise identical 2D control runs where one species is compressed uniformly in all directions and remains nearly isotropic, mirror growth is not observed. Both ions and electrons exhibit nonthermal tails in their energy spectra, as expected from stochastic acceleration mediated by scattering off their respective cyclotron wave fluctuations \citep{riquelme_stochastic_electron_acceleration_2017, ley_stochastic_ion_acceleration_2019}.

    
    \item To evaluate the sensitivity of plasma instabilities to magnetization and thermal content, we scan $\beta_{i0}$ over a factor of 20 and $\Theta_{i0}$ over a factor of 8. Figure \ref{fig:simulation_suite_late_time_anisotropy_perturbations} summarizes the late-time results, showing the deviation of anisotropy from the adiabatic baseline at $t=2\,q^{-1}$ and the normalized magnetic field perturbations over $t=1.5$--$2\,q^{-1}$. We omit cases where the species follow adiabatic growth for the entire duration of the simulation. A larger deviation from the adiabatic prediction indicates a lower anisotropy threshold. At fixed temperature, the deviation increases with $\beta_{i0}$. As the plasma becomes increasingly relativistic (higher $\Theta_{i0}$), the anisotropy threshold for all instabilities increases, leading to smaller deviations from the adiabatic prediction (at fixed $\beta_{i0}$). This is consistent with relativistic linear theory for whistler modes \citep{gladd_relativistic_whistler_1983}, and we expect a similar trend for ion cyclotron waves given the cyclotron nature of both modes. Furthermore, the amplitudes of the cyclotron-wave (panel (c)) and mirror-mode (panel (d)) fluctuations increase with increasing $\beta_{i0}$ and decreasing $\Theta_{i0}$. For $\beta_{i0}<0.5$, the mirror threshold is sufficiently high that the mirror instability does not develop for any $\Theta_{i0}$ over the duration of our simulations (till $t=2\,q^{-1}$). 
    
    \item Low-collisionality accretion flows are expected to be two-temperature with $T_e<T_i$ due to inefficient coupling between electrons and ions. Due to computational constraints, we restrict the 2D $T_{e0}/T_{i0}$ survey to $1$ and $0.5$. Lowering the electron temperature delays the onset of the mirror instability, consistent with linear theory indicating reduced mirror growth rates at lower $T_e$ when $T_{\perp,e}>T_{\parallel,e}$ \citep{pokhotelov_electron_anisotropy_mirror_2000}. For an initially colder electron population, the saturated electron anisotropy $A_e$ is higher, a trend we confirm down to $T_{e0}/T_{i0}=0.25$ with 1D simulations. For $T_{e0}/T_{i0}<0.5$, holding all other parameters at their fiducial values, the electrons remain consistent with adiabatic evolution over the simulated interval (until $t=2\,q^{-1}$). This suggests that, absent other effects, e.g., radiative cooling, adiabatic evolution of electron pressure anisotropy provides a simple baseline for global models. We also find that colder electrons develop a weaker nonthermal tail, a result we verify with 1D simulations in which the ions are compressed isotropically to isolate the effect of electron-only physics.
    
    \item We find that slower compression (larger $\omega_{c,i0}/q$) leads to an earlier onset of all instabilities in units of $q^{-1}$. Differences in the onset times of the ion cyclotron and mirror instabilities diminish with increasing $\omega_{c,i0}/q$. We confirm this trend for ion cyclotron waves using 1D runs, which suppress the growth of mirror modes, carried out at even slower compression rates, up to $\omega_{c,i0}/q=12800$. Faster compression produces a larger overshoot of $A_i$ above the ion cyclotron anisotropy threshold and yields larger saturated values of $(\delta B_{\perp}/B_0)^2$. Nevertheless, the late-time ($t=2\,q^{-1}$) value of $A_i$ shows little dependence on the compression rate. Similarly, the saturated electron anisotropy shows only a modest dependence on $\omega_{c,i0}/q$. The mirror instability saturates at comparable $(\delta B_{\parallel}/B_0)^2$ across the range of $\omega_{c,i0}/q$ explored.
\end{enumerate}

In this work, we have considered a relatively short compression timescale compared with the expected timescale of large-scale motions in accretion disks---e.g., shear, compression, and expansion---which are comparable to the dynamical time. In accretion disks, the compression rate is extremely slow in gyrofrequency units $\omega_{c,i}/q\sim10^8$, indicating a vast separation between global and kinetic timescales. The onset times of the various instabilities appear to converge as $\omega_{c,i0}/q$ increases (Section \ref{sec:results_compression_rate}), but over the range of compression rates probed here it remains unclear what asymptotic onset time will be reached in the slow-driving limit. Extending this study to slower compressions is computationally demanding, requiring not only more timesteps to reach the same final time in units of $q^{-1}$, but also a higher \texttt{ppc} to suppress the numerical noise that accumulates prior to the onset of the instabilities.


The simulations presented here are 2D, complemented by selected 1D runs. Some of our results might change in 3D, most notably for the mirror mode, which is unstable over a wider range of wavevectors in 3D, where an additional perpendicular direction becomes available, and may therefore carry more magnetic energy density relative to the ion cyclotron waves than in 2D \citep{shoji_mirror_3d_2009, markovskii_mirror_3d_humps_holes_2025}. Its role in regulating the ion anisotropy could thus be correspondingly larger in 3D. We leave a dedicated study of 3D effects to future work.

The compressing-box setup is meant to represent a fluid element advected through an accretion flow. While we focus here on compression---guided by the dominant $P_{\perp}>P_{\parallel}$ state observed in global weakly collisional MAD models---the global simulations also develop localized firehose-prone patches with $P_{\perp}<P_{\parallel}$, which can influence electromagnetic observables \citep{galishnikova_anisotropic_images_2023}. We defer a dedicated exploration of this regime to future work. Additionally, although we consider here a simple mechanism for generating particle anisotropy, namely magnetic field amplification coupled with conservation of adiabatic invariants, realistic black hole accretion environments involve a variety of processes capable of generating pitch-angle anisotropy, such as magnetic reconnection \citep{comisso_pitch_angle_reconnection_pair_plasma_2024, comisso_pitch_angle_reconnection_ion_electron_2024}, turbulent cascade \citep{comisso_ion_electron_acceleration_turbulence_2022}, and synchrotron cooling (in radiatively efficient systems; \citealp{zhdankin_sfhi_2023}). Also, our simulations neglect radiative losses, an approximation justified for radiatively inefficient flows such as that around \sgra, where the synchrotron cooling time of the bulk electrons greatly exceeds the dynamical time. More generally, since radiative cooling acts most rapidly on the highest-energy particles, it can affect the nonthermal tails even when the bulk plasma is unaffected.

We have restricted our analysis to an electron-proton plasma, neglecting the heavier nuclei likely present in the Galactic Center accretion flow, which is fed by stellar winds \citep{najarro_spectroscopy_galactic_center_1997, paumard_galactic_center_star_disks_2006}. Composition effects can affect inferences from Sgr A* when confronting models with synthetic observables \citep{wong_helium_survey_2022}. A natural extension of this work is to consider how heavier ions can modify the mirror and ion cyclotron instability thresholds \citep{price_mirror_ion_cyclotron_helium_1986,gary_ion_anisotropy_magnetosheath_helium_1993,gary_ion_cyclotron_terrestrial_magnetosheath_helium_1993,remya_electron_anisotropy_ic_mirror_2013}.

Lastly, we have focused here on the time evolution and spectral properties of the dominant plasma instabilities in a low-$\beta$, transrelativistic plasma, and on their role in regulating ion and electron pressure anisotropy, with a limited discussion of how these fluctuations shape the particle energy spectra. In a future study, we will quantify particle energization in this regime and identify the channels responsible for irreversible heating and accompanying nonthermal acceleration.

Note: After this manuscript was submitted, we became aware of \citet{ley_relativistic_cgl_2026, wierzchucka_relativistic_cgl_2026}, who extend the double-adiabatic equation of state of \citet{chew_cgl_equation_of_state_1956} to the relativistic regime, thereby obtaining evolution equations for $P_{\perp}$ and $P_{\parallel}$. In Appendix~\ref{app:cgl_comparison} we show that the evolution of the electron temperature (equivalently, pressure) anisotropy in our simulations follows the predicted behavior in the ultrarelativistic regime ($\Theta_e \gg 1$). We further note that repeating the analysis of Figure~\ref{fig:simulation_suite_late_time_anisotropy_perturbations}(b) with the temperature-based anisotropy ($A_{\text{T,e}}\equiv T_{\perp,e}/T_{\parallel,e}-1$) and the relativistic CGL-like baseline yields values of $\Delta_e$ nearly indistinguishable from those shown.



\begin{acknowledgements}
The authors thank Charles Gammie, Kristopher Klein and Daniel Verscharen for insightful discussions throughout this work. V.D. is supported by the Dissertation Completion Fellowship and the Donald C. and F. Shirley Jones Fellowship. This research is part of the Delta research computing project, which is supported by the National Science Foundation (award OCI 2005572), and the State of Illinois. Delta is a joint effort of the University of Illinois at Urbana-Champaign and its National Center for Supercomputing Applications. L.S. was supported by NSF grant PHY2409223. The work was supported by a grant from the Simons Foundation (MP-SCMPS-0000147, to L.S.). L.S. also acknowledges support from the Department of Energy (DOE) Early Career Award DE-SC0023015. A.T. was supported by NSF PHY2010189 and the DOE Fusion Energy Sciences Postdoctoral Research Program, administered by the Oak Ridge Institute for Science and Education and Oak Ridge Associated Universities under DOE contract DE-SC0014664. The data analysis was possible thanks to the high-throughput computing utility `Launcher' \citep{wilson_launcher_2017}.
\end{acknowledgements}

\appendix

\section{Transient Longitudinal Perturbations during Ion Cyclotron Growth}\label{appendix:dby_bump}

\begin{figure*}
\centering
\includegraphics[width=\textwidth]{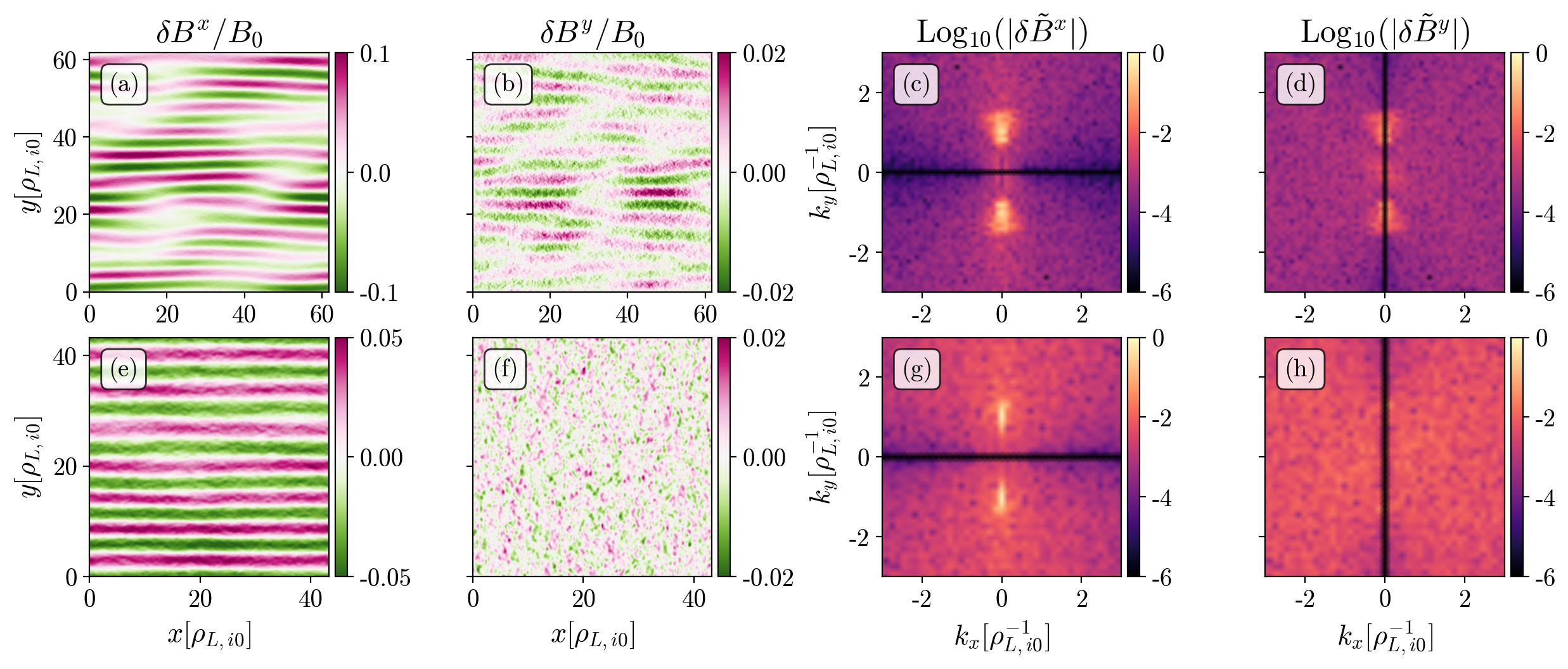}
\caption{Transverse ($\delta B^x$) and longitudinal ($\delta B^y$) magnetic field perturbations for two simulations. We examine these profiles to identify the origin of the transient longitudinal fluctuations that appear near ion-cyclotron saturation in several simulations. Rows 1 and 2 display snapshots of field perturbations for Simulation 1 ($T_{e0}/T_{i0}=1,\,\beta_{i0}=0.2,\,\Theta_{i0}=0.025$) and Simulation 2 ($T_{e0}/T_{i0}=1,\,\beta_{i0}=0.5,\,\Theta_{i0}=0.1$), respectively. The left two columns show the field perturbations $\delta B^x$ and $\delta B^y$ in real space, and the right two columns show the same field perturbations in Fourier space $(k_x, k_y)$. The spatial spectra are normalized by the global maximum across both components. Consequently, the peak power in $\delta\tilde{B}^y$ appears significantly smaller than in $\delta\tilde{B}^x$, reflecting its subdominant nature. The zero power horizontal (vertical) lines in $\delta\tilde{B}^x$ ($\delta\tilde{B}^y$) are a consequence of mean subtraction and $\boldsymbol{\nabla\cdot B}=0$.} 
\label{fig:bump_longitudinal_perturbations_during_ic_growth}
\end{figure*}

In several simulations presented in this study, we observe a transient feature in the longitudinal magnetic field perturbations $\delta B^y$ at the end of the linear growth phase of the ion cyclotron instability. This manifests as a distinct bump in the time series and appears as low-amplitude striations (relative to the transverse perturbations) in $\delta B^y$ in the $(x,y)$ plane. This can be understood as a result of the ion cyclotron modes developing a finite obliquity relative to the mean field $\boldsymbol{B}_0$. Any deviation from strictly parallel propagation in 2D $(x,y)$ requires longitudinal perturbations to ensure the magnetic field remains divergence-free $i\boldsymbol{k}\cdot\delta\tilde{\boldsymbol{B}}=0$.

Figure \ref{fig:bump_longitudinal_perturbations_during_ic_growth} compares magnetic field perturbations from two representative simulations at the end of the linear ion cyclotron phase: one that develops the transient longitudinal feature (top row) and one that does not (bottom row). For the case where the ion cyclotron modes develop obliquity with respect to the mean field $\boldsymbol{B}_0$, $\delta\tilde{B}^x$ shows finite spread in $k_x$ (panel c) and $\delta\tilde{B}^y$ develops subdominant power at the same $k_y$ (panel d), consistent with the striations in $\delta B^y$ (panel b). In the case where the ion cyclotron modes are field aligned, the spectrum is confined to $k_x\simeq 0$ (panel g) and $\delta\tilde{B}^y$ shows no coherent structure (panel h).
\section{Numerical Convergence}\label{appendix:convergence}

\begin{figure*}
\centering
\includegraphics[width=\textwidth]{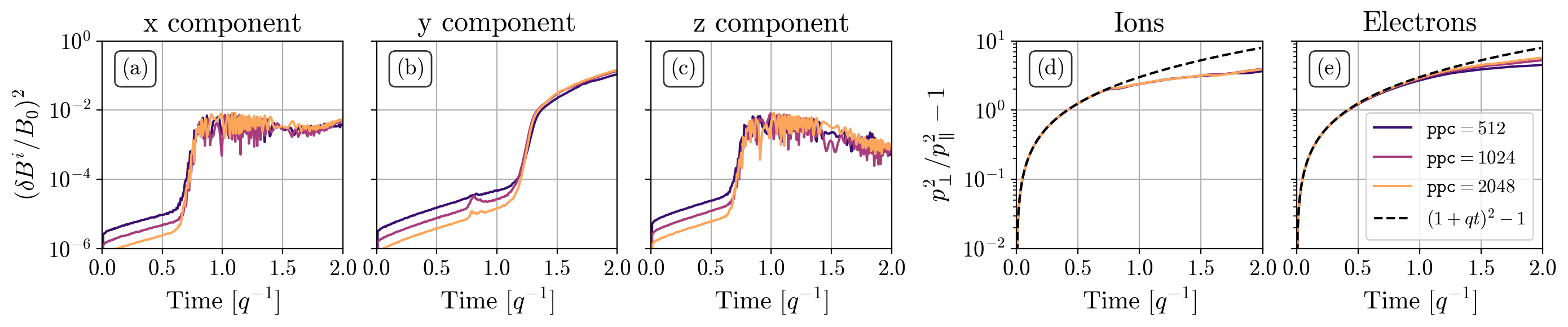}
\caption{Convergence with respect to the initial \texttt{ppc} for the fiducial 2D simulation ($T_{e0}/T_{i0}=1,\,\beta_{i0}=0.5$ and $\Theta_{i0}=0.05$). Left three columns: Time series of perturbations in the magnetic field normalized by the mean field. Right two columns: Time evolution of the ion and electron anisotropy.} 
\label{fig:convergence_fiducial_ppc}
\end{figure*}

\begin{figure}
\centering
\includegraphics[width=0.47\textwidth]{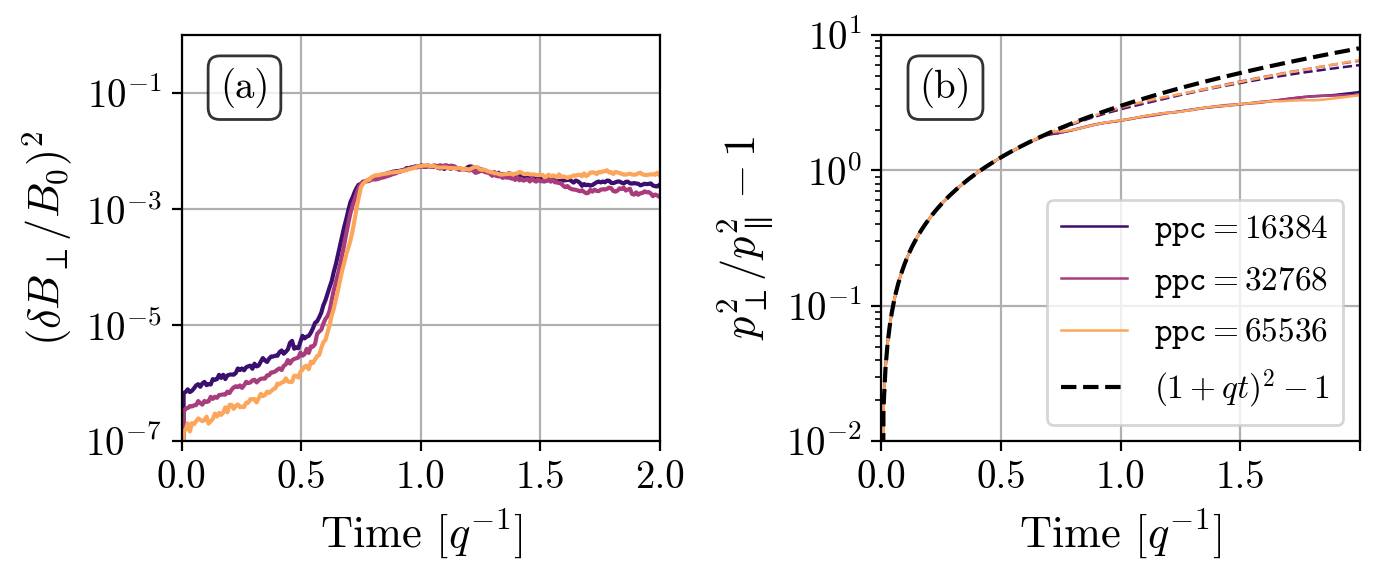}
\caption{Convergence with respect to the initial \texttt{ppc} for the fiducial 1D simulation. Panel (a): Time series of the perpendicular component of the magnetic field. Panel (b): Time evolution of the ion (solid) and electron (dashed) anisotropy.} 
\label{fig:convergence_fiducial_1d_ppc}
\end{figure}

\begin{figure*}
\centering
\includegraphics[width=\textwidth]{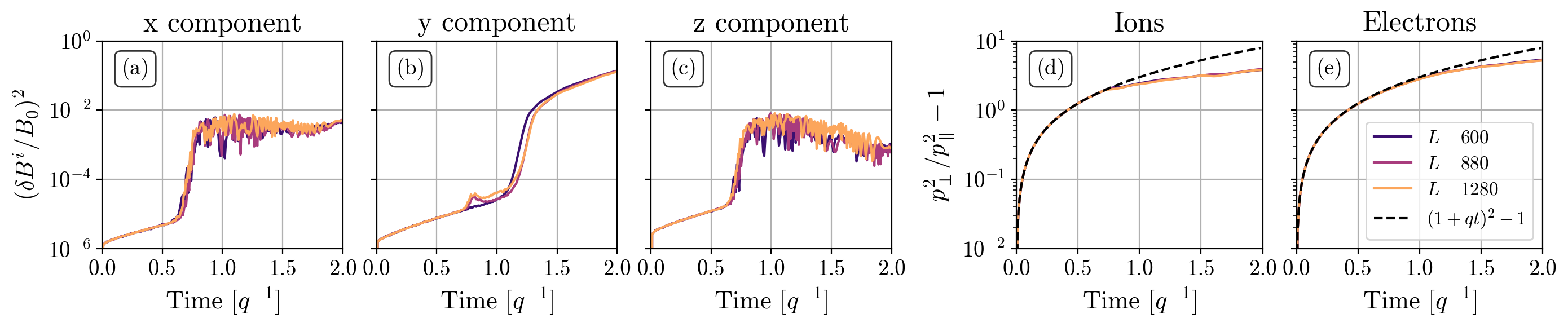}
\caption{Similar to Figure \ref{fig:convergence_fiducial_ppc} but here we show convergence with respect to the box size for the fiducial 2D simulation ($T_{e0}/T_{i0}=1,\,\beta_{i0}=0.5$ and $\Theta_{i0}=0.05$). $L$ denotes the total number of cells along each side of the box. The initial electron skin depth is the same for all cases considered in this convergence study.} 
\label{fig:convergence_fiducial_box_size}
\end{figure*}

\begin{figure*}
\centering
\includegraphics[width=\textwidth]{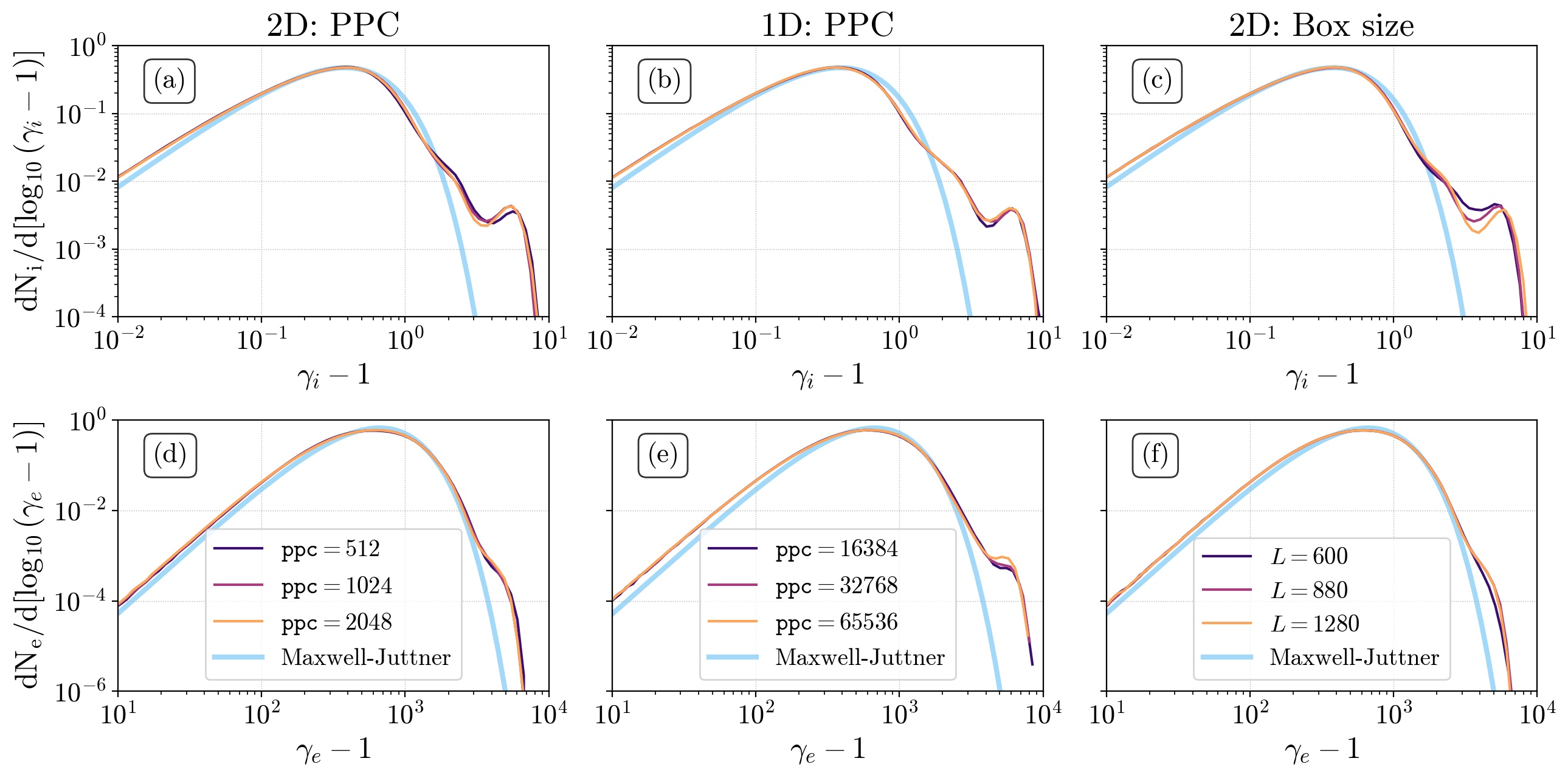}
\caption{Convergence of the ion (top row) and electron (bottom row) energy spectra at $t\simeq2\,q^{-1}$, for the fiducial parameters. Panels (a) and (d) show the 2D simulations initialized with $\mathtt{ppc}=512,1024,2048$; panels (b) and (e) the corresponding 1D simulations with $\mathtt{ppc}=16384,32768,65536$; and panels (c) and (f) the 2D simulations with box sizes $L=600,880,1280$. The thick light blue curves show the isotropic Maxwell--J\"uttner distributions. Each column corresponds to the scan shown in Figures~\ref{fig:convergence_fiducial_ppc}, \ref{fig:convergence_fiducial_1d_ppc}, and \ref{fig:convergence_fiducial_box_size}, respectively.}
\label{fig:convergence_fiducial_spectra}
\end{figure*}

For our fiducial simulation, we now present convergence results with respect to the initial number of particles per cell and the size of the compressing box.

Figure \ref{fig:convergence_fiducial_ppc} shows the evolution of the magnetic field energy and the ion and electron anisotropy for 2D simulations initialized with a different number of particles per cell, $\mathtt{ppc}=512,1024,2048$. Since our focus is on capturing both ion- and electron-scale physics in black hole accretion flows, ensuring convergence in electron dynamics is essential. We find that the relative difference in the saturated electron anisotropy (as measured at $t = 2\,q^{-1}$) between the fiducial \texttt{ppc} and the simulation with twice the \texttt{ppc} is less than 6\%. We consider this level of agreement to be a satisfactory criterion for convergence. We also find convergence for our 1D simulations as highlighted in Figure \ref{fig:convergence_fiducial_1d_ppc}.

To ensure that our simulations resolve at least a few wavelengths of the dominant ion- and electron-driven instabilities, we perform a convergence test by varying the size of the compressing box. As shown in Figure~\ref{fig:convergence_fiducial_box_size}, we find excellent agreement between the fiducial simulation and one with a box that is $\sim1.5$x larger in each spatial dimension. In all simulations, we fix $c/\omega_{p,e0}=10$, thereby effectively probing the dependence of our results on the ratio $L/(c/\omega_{p,e0})$.

As shown in Figure~\ref{fig:convergence_fiducial_spectra}, the ion and electron energy spectra, including the nonthermal tails, are likewise insensitive to \texttt{ppc} and box size. The only appreciable difference is in the depth of the trough separating the thermal peak from the suprathermal bump in the ion spectrum (panel (c)), which is slightly shallower in the larger boxes, while the location and amplitude of the bump itself, the slope of the tail, and the maximum energy attained are unchanged. Since our discussion of the spectra concerns the presence of the nonthermal component and its qualitative dependence on plasma parameters rather than a precise characterization of the spectral shape, we consider this level of agreement sufficient for the conclusions drawn here.
\section{Comparison with relativistic CGL prediction}\label{app:cgl_comparison}
\begin{figure}
\centering
\includegraphics[,width=0.9\linewidth]{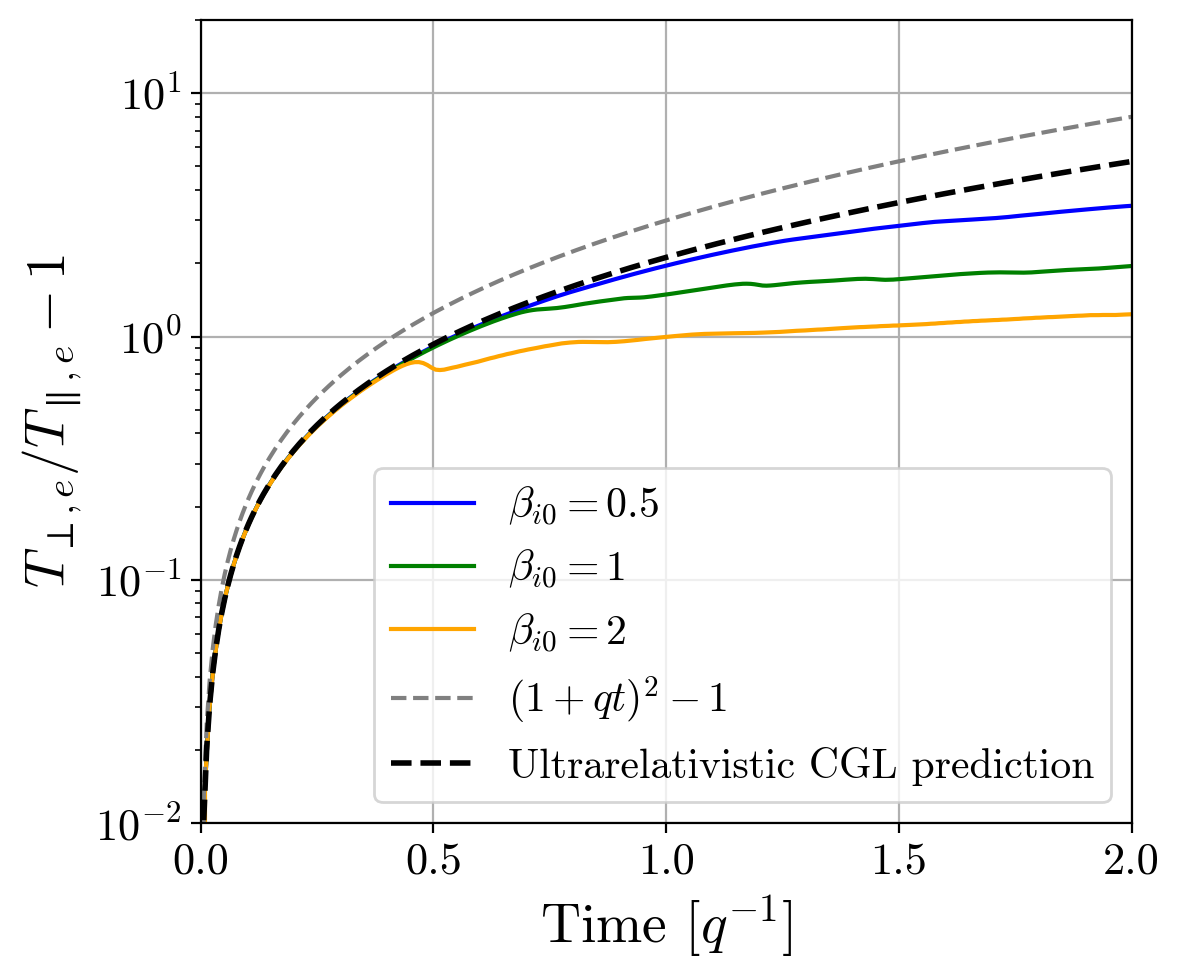}
\caption{Time series of the electron temperature anisotropy $A_{\mathrm{T},e} \equiv T_{\perp,e}/T_{\parallel,e} - 1$ for a representative sample of simulations with $T_{e0}/T_{i0}=1$ and $\Theta_{i0}=0.05$, at varying $\beta_{i0}$ (colored solid lines; blue corresponds to the fiducial simulation). The gray dashed line shows the nonrelativistic double-adiabatic prediction, $(1+qt)^2 - 1$, while the black dashed line shows the ultrarelativistic prediction for a compressing box \citep{ley_relativistic_cgl_2026, wierzchucka_relativistic_cgl_2026}. The simulated anisotropy follows the ultrarelativistic prediction until instabilities set in and drive the anisotropy toward the marginal stability threshold.}
\label{fig:electron_anisotropy_relativistic_cgl_comparison}
\end{figure}

The standard double-adiabatic model of \citet{chew_cgl_equation_of_state_1956} assumes nonrelativistic particles and therefore does not correctly capture the temperature anisotropy evolution when the plasma is relativistically hot ($\Theta \gtrsim 1$), as is the case for the electrons in our simulations. \citet{ley_relativistic_cgl_2026} and  \citet{wierzchucka_relativistic_cgl_2026} have recently extended these equations to the relativistic regime, and in particular provide their ultrarelativistic limit. For the case of a compressing box, Equations~4.4 and~4.6 of \citet{ley_relativistic_cgl_2026} give the evolution of $P_{\perp}$ and $P_{\parallel}$, respectively, under conservation of the adiabatic invariants. Figure~\ref{fig:electron_anisotropy_relativistic_cgl_comparison} compares the simulated electron temperature anisotropy $A_{\mathrm{T},e}$ against this prediction (black dashed) and the nonrelativistic double-adiabatic expectation (gray dashed), for a representative subset of runs with $T_{e0}/T_{i0}=1$ and $\Theta_{i0}=0.05$ spanning $\beta_{i0} \in \{0.5, 1, 2\}$. At early times, all simulations track the ultrarelativistic CGL prediction closely and depart substantially from the nonrelativistic curve, confirming that the ultrarelativistic treatment is the appropriate baseline for the electrons. The subsequent departure from the black dashed curve reflects the onset of kinetic instabilities that regulate the anisotropy toward the marginal stability threshold, with larger $\beta_{i0}$ leading to earlier and stronger deviation---consistent with the trends discussed in the main text.

\section{Characteristic Length and Frequency Scales in a Relativistic Plasma}\label{appendix:scales}

In this work, we study the evolution of a relativistic electron–ion plasma, with a realistic mass ratio, in a compressing box. To capture the kinetic physics of both species, the simulations must resolve the relevant characteristic length and time scales, which in the relativistic regime require accounting for the particles’ relativistic inertia. Here, we express these scales in terms of the input parameters $T_{e0}/T_{i0}$, $\Theta_{i0}$, $c/\omega_{p,e0}$, $m_i/m_e$, $\omega_{c,i0}/q$, and the initial magnetization $\sigma_0$ (determined from $\beta_{i0}$ and other parameters).

The initial magnetic field $B_0$ then follows from $\sigma_0 \equiv B_0^2/(4\pi w)$,
where $w = (n_0 m_{i} + n_0 m_{e}) c^2 + \Gamma_{i} u_i + \Gamma_{e} u_e$ is the relativistic enthalpy density. Here, $n_0$ is the initial number density of each species, and $\Gamma_j$ and $u_j$ are the adiabatic index and internal energy density of species $j$ (ions or electrons), respectively. The adiabatic indices are calculated using a fitting formula for a perfect, relativistic, Boltzmann gas \citep{service_fitting_formula_synge_1986},
\begin{align}
    \Gamma_j = \frac{1}{3}\Big(5 &- 1.21937z + 0.18203z^2 - 0.96583z^3 \nonumber \\
    &+ 2.32513z^4 - 2.39332z^5 + 1.07136z^6\Big),
\end{align}
where $z\equiv\Theta_j/(0.24+\Theta_j)$. Using the relation for internal energy density, $u_j=n_{j}m_j c^2\Theta_j/(\Gamma_j-1)$, the total enthalpy density becomes,
\begin{equation}
    w = n_{i}m_i c^2\bigg[1 + \frac{\Gamma_i\Theta_{i0}}{\Gamma_i-1} + \frac{m_e}{m_i}\Big(1 + \frac{\Gamma_e\Theta_{e0}}{\Gamma_e-1}\Big)\Bigg],
\end{equation}
where the initial electron temperature $\Theta_{e0}$ is related to the ion temperature by $\Theta_{e0} = \Theta_{i0}\,(T_{e0}/T_{i0})\,(m_i/m_e)$.

The initial ion plasma frequency, $\omega_{p,i0}$ expressed in terms of the initial electron skin depth, $c/\omega_{p,e0}$ is,
\begin{equation}
    \omega_{p,i0} = \frac{c}{c/\omega_{p,e0}}\cdot\sqrt{\frac{\langle\gamma_e\rangle m_e}{\langle\gamma_i\rangle m_i}},
\end{equation}
where the average Lorentz factor for species $j$ is $\langle\gamma_j\rangle=1+\Theta_{j0}/(\Gamma_j-1)$. The compression rate is defined in units of the inverse ion cyclotron frequency, which can be written in terms of the input parameters as,
\begin{align}
    \omega_{c,i0} = \frac{c}{c/\omega_{p,e0}}&\cdot\sqrt{\frac{\langle\gamma_e\rangle m_e}{\langle\gamma_i\rangle m_i}}\cdot\sqrt{\frac{\sigma_0}{\langle\gamma_i\rangle}} \nonumber \\
    &\cdot\bigg[1 + \frac{\Gamma_i\Theta_{i0}}{\Gamma_i-1} + \frac{m_e}{m_i}\Big(1 + \frac{\Gamma_e\Theta_{e0}}{\Gamma_e-1}\Big)\bigg]^{1/2},
\end{align}
and the corresponding electron cyclotron frequency is,
\begin{align}
    \omega_{c,e0} = \frac{c}{c/\omega_{p,e0}}&\cdot\sqrt{\frac{m_i}{\langle\gamma_e\rangle m_e}}\cdot\sqrt{\sigma_0} \nonumber \\
    &\cdot\bigg[1 + \frac{\Gamma_i\Theta_{i0}}{\Gamma_i-1} + \frac{m_e}{m_i}\Big(1 + \frac{\Gamma_e\Theta_{e0}}{\Gamma_e-1}\Big)\bigg]^{1/2}.
\end{align}
The initial ion gyroradius, $\rho_{L,i0}$ is given by,
\begin{align}
    \rho_{L,i0} = \frac{c}{\omega_{p,e0}}&\cdot\sqrt{\frac{m_i}{m_e}}\cdot\sqrt{\frac{1}{\langle\gamma_e\rangle}}\cdot\sqrt{\frac{1}{\sigma_0}}\cdot\langle\gamma_i v_{\perp,i0}\rangle \nonumber \\
    &\cdot\bigg[1 + \frac{\Gamma_i\Theta_{i0}}{\Gamma_i-1} + \frac{m_e}{m_i}\Big(1 + \frac{\Gamma_e\Theta_{e0}}{\Gamma_e-1}\Big)\bigg]^{-1/2},
\end{align}
where, recall that $v_{\perp,i0}$ is the initial ion perpendicular velocity. Both species are initialized with an isotropic Maxwell-J\"{u}ttner distribution, giving $\langle\gamma_i v_{\perp,i0}\rangle/c=\pi/2\cdot e^{-1/\Theta_{i0}}\,\Theta_{i0}\,(1+3\Theta_{i0}+3\Theta_{i0}^2)/K_2(\Theta_{i0}^{-1})$, where $K_2$ is the modified Bessel function of the second kind. Similarly, the initial electron gyroradius $\rho_{L,e0}\equiv\langle\gamma_e v_{\perp,e0}\rangle m_e c / (\vert q_e\vert B_0)$ is,
\begin{align}
    \rho_{L,e0} = \frac{c}{\omega_{p,e0}}&\cdot\sqrt{\frac{m_e}{m_i}}\sqrt{\frac{1}{\langle\gamma_e\rangle}}\cdot\sqrt{\frac{1}{\sigma_0}}\cdot\langle\gamma_e v_{\perp,e0}\rangle \nonumber \\
    &\cdot\bigg[1 + \frac{\Gamma_i\Theta_{i0}}{\Gamma_i-1} + \frac{m_e}{m_i}\Big(1 + \frac{\Gamma_e\Theta_{e0}}{\Gamma_e-1}\Big)\bigg]^{-1/2},
\end{align}
where $\langle\gamma_e v_{\perp,e0}\rangle/c = \pi/2\cdot e^{-1/\Theta_{e0}}\,\Theta_{e0}\,(1+3\Theta_{e0}+3\Theta_{e0}^2)/K_2(\Theta_{e0}^{-1})$

\section{Simulation parameters}\label{appendix:simulation_parameters}

Table \ref{table:simulation_input_parameters} summarizes the input parameters for the simulations discussed in the main text. We omit runs performed exclusively for numerical convergence (Appendix \ref{appendix:convergence}). If either species is compressed isotropically, we choose the compression rate $q_{\text{iso}} = 2q/3$. The parameters are defined as follows:\begin{enumerate}
    \item $\beta_{i0}$ is the ion plasma $\beta$.
    \item $\Theta_{i0}$ is the ion temperature in units of its rest mass energy.
    \item $T_{e0}/T_{i0}$ is the electron-to-ion temperature ratio.
    \item $v_{A,i0}/c$ is the ion Alfv\'{e}n speed in units of the speed of light.
    \item $\omega_{c,i0}/q$ is the inverse compression rate expressed in units of the ion cyclotron frequency.
    \item $L$ denotes the domain size, measured in number of cells. In 2D simulations, the domain is square, with $L_x=L_y=L$. In 1D simulations, $L_y=L$ and $L_x=2$ cells. 
    \item $\texttt{c\_omp}$ is the electron skin depth ($c/\omega_{p,e0}$) measured in number of cells.
    \item $\texttt{ppc}$ is the total number of particles per cell.
    \item $\texttt{isoion}$ indicates a run in which the ions are compressed isotropically.
    \item $\texttt{isolec}$ indicates a run in which the electrons are compressed isotropically.
    \item `Dimensions' indicates the dimensionality of the run (1D vs 2D).
\end{enumerate}

\setlength{\tabcolsep}{4.5pt}
\begin{deluxetable*}{ lccccccccccc }
\tablecaption{Simulation suite} \label{table:simulation_input_parameters}
\tablehead{
\colhead{Purpose} & 
\colhead{$\beta_{i0}$} &
\colhead{$\Theta_{i0}$} &
\colhead{$T_{e0}/T_{i0}$} &
\colhead{$v_{A,i0}/c$} &
\colhead{$\omega_{c,i0}/q$} &
\colhead{$L$} &
\colhead{\texttt{c\_omp}} &
\colhead{\texttt{ppc}} &
\colhead{\texttt{isoion}} &
\colhead{\texttt{isolec}} &
\colhead{Dimensions}
}
\startdata
Fiducial case & 0.5 & 0.05 & 1 & 0.408 & 1600 & 880 & 10 & 1024 & 0 & 0 & 2D \\
Fiducial case & 0.5 & 0.05 & 1 & 0.408 & 1600 & 880 & 10 & 1024 & 1 & 0 & 2D \\
Fiducial case & 0.5 & 0.05 & 1 & 0.408 & 1600 & 880 & 10 & 1024 & 0 & 1 & 2D \\
Fiducial case & 0.5 & 0.05 & 1 & 0.408 & 1600 & 1200 & 10 & 32768 & 0 & 0 & 1D \\
\hline
Parameter survey ($\beta_{i0}$ , $\Theta_{i0}$) & 0.1 & 0.025 & 1 & 0.577 & 1600 & 800 & 10 & 1024 & 0 & 0 & 2D \\
Parameter survey ($\beta_{i0}$ , $\Theta_{i0}$) & 0.2 & 0.025 & 1 & 0.447 & 1600 & 1200 & 10 & 2048 & 0 & 0 & 2D \\
Parameter survey ($\beta_{i0}$ , $\Theta_{i0}$) & 0.5 & 0.025 & 1 & 0.302 & 1600 & 1200 & 10 & 2048 & 0 & 0 & 2D \\
Parameter survey ($\beta_{i0}$ , $\Theta_{i0}$) & 1 & 0.025 & 1 & 0.218 & 1600 & 880 & 6 & 1024 & 0 & 0 & 2D \\
Parameter survey ($\beta_{i0}$ , $\Theta_{i0}$) & 2 & 0.025 & 1 & 0.156 & 1600 & 760 & 6 & 1024 & 0 & 0 & 2D \\
Parameter survey ($\beta_{i0}$ , $\Theta_{i0}$) & 0.1 & 0.05 & 1 & 0.707 & 1600 & 950 & 12 & 1024 & 0 & 0 & 2D \\
Parameter survey ($\beta_{i0}$ , $\Theta_{i0}$) & 0.2 & 0.05 & 1 & 0.577 & 1600 & 880 & 10 & 1024 & 0 & 0 & 2D \\
Parameter survey ($\beta_{i0}$ , $\Theta_{i0}$) & 1 & 0.05 & 1 & 0.302 & 1600 & 800 & 10 & 1024 & 0 & 0 & 2D \\
Parameter survey ($\beta_{i0}$ , $\Theta_{i0}$) & 1 & 0.05 & 1 & 0.302 & 1600 & 800 & 10 & 1024 & 1 & 0 & 2D \\
Parameter survey ($\beta_{i0}$ , $\Theta_{i0}$) & 2 & 0.05 & 1 & 0.218 & 1600 & 880 & 10 & 1024 & 0 & 0 & 2D \\
Parameter survey ($\beta_{i0}$ , $\Theta_{i0}$) & 2 & 0.05 & 1 & 0.302 & 1600 & 880 & 10 & 1024 & 1 & 0 & 2D \\
Parameter survey ($\beta_{i0}$ , $\Theta_{i0}$) & 0.1 & 0.1 & 1 & 0.816 & 1600 & 880 & 16 & 1024 & 0 & 0 & 2D \\
Parameter survey ($\beta_{i0}$ , $\Theta_{i0}$) & 0.2 & 0.1 & 1 & 0.707 & 1600 & 800 & 12 & 1024 & 0 & 0 & 2D \\
Parameter survey ($\beta_{i0}$ , $\Theta_{i0}$) & 0.5 & 0.1 & 1 & 0.534 & 1600 & 600 & 10 & 1024 & 0 & 0 & 2D \\
Parameter survey ($\beta_{i0}$ , $\Theta_{i0}$) & 1 & 0.1 & 1 & 0.408 & 1600 & 600 & 10 & 1024 & 0 & 0 & 2D \\
Parameter survey ($\beta_{i0}$ , $\Theta_{i0}$) & 2 & 0.1 & 1 & 0.302 & 1600 & 600 & 10 & 1024 & 0 & 0 & 2D \\
Parameter survey ($\beta_{i0}$ , $\Theta_{i0}$) & 2 & 0.1 & 1 & 0.302 & 1600 & 600 & 10 & 1024 & 1 & 0 & 2D \\
Parameter survey ($\beta_{i0}$ , $\Theta_{i0}$) & 0.1 & 0.2 & 1 & 0.894 & 1600 & 1200 & 30 & 1024 & 0 & 0 & 2D \\
Parameter survey ($\beta_{i0}$ , $\Theta_{i0}$) & 0.2 & 0.2 & 1 & 0.816 & 1600 & 1200 & 30 & 1024 & 0 & 0 & 2D \\
Parameter survey ($\beta_{i0}$ , $\Theta_{i0}$) & 0.5 & 0.2 & 1 & 0.667 & 1600 & 800 & 20 & 1024 & 0 & 0 & 2D \\
Parameter survey ($\beta_{i0}$ , $\Theta_{i0}$) & 1 & 0.2 & 1 & 0.534 & 1600 & 600 & 10 & 2048 & 0 & 0 & 2D \\
Parameter survey ($\beta_{i0}$ , $\Theta_{i0}$) & 2 & 0.2 & 1 & 0.408 & 1600 & 600 & 10 & 1024 & 0 & 0 & 2D \\
Parameter survey ($\beta_{i0}$ , $\Theta_{i0}$) & 2 & 0.2 & 1 & 0.408 & 1600 & 600 & 10 & 1024 & 1 & 0 & 2D \\
\hline
Parameter survey ($T_{e0}/T_{i0}$) & 0.5 & 0.05 & 0.5 & 0.408 & 1600 & 880 & 7 & 2048 & 0 & 0 & 2D \\
Parameter survey ($T_{e0}/T_{i0}$) & 1 & 0.05 & 0.5 & 0.302 & 1600 & 880 & 5 & 2048 & 0 & 0 & 2D \\
Parameter survey ($T_{e0}/T_{i0}$) & 2 & 0.1 & 0.5 & 0.302 & 1600 & 1024 & 10 & 1024 & 0 & 0 & 2D \\
Parameter survey ($T_{e0}/T_{i0}$) & 0.5 & 0.05 & 0.5 & 0.408 & 1600 & 880 & 7 & 65536 & 0 & 0 & 1D \\
Parameter survey ($T_{e0}/T_{i0}$) & 0.5 & 0.05 & 0.25 & 0.408 & 1600 & 1760 & 10 & 65536 & 0 & 0 & 1D \\
Parameter survey ($T_{e0}/T_{i0}$) & 0.5 & 0.05 & 1 & 0.408 & 1600 & 800 & 10 & 32768 & 1 & 0 & 1D \\
Parameter survey ($T_{e0}/T_{i0}$) & 0.5 & 0.05 & 0.75 & 0.408 & 1600 & 1000 & 10 & 65536 & 1 & 0 & 1D \\
Parameter survey ($T_{e0}/T_{i0}$) & 0.5 & 0.05 & 0.5 & 0.408 & 1600 & 880 & 7 & 65536 & 1 & 0 & 1D \\
\hline
Parameter survey ($\omega_{c,i0}/q$) & 0.5 & 0.05 & 1 & 0.408 & 800 & 880 & 10 & 1024 & 0 & 0 & 2D \\
Parameter survey ($\omega_{c,i0}/q$) & 0.5 & 0.05 & 1 & 0.408 & 3200 & 880 & 10 & 2048 & 0 & 0 & 2D \\
Parameter survey ($\omega_{c,i0}/q$) & 0.5 & 0.05 & 1 & 0.408 & 6400 & 880 & 10 & 1024 & 0 & 0 & 2D \\
Parameter survey ($\omega_{c,i0}/q$) & 0.5 & 0.05 & 1 & 0.408 & 800 & 1200 & 10 & 65536 & 0 & 0 & 1D \\
Parameter survey ($\omega_{c,i0}/q$) & 0.5 & 0.05 & 1 & 0.408 & 3200 & 1200 & 10 & 65536 & 0 & 0 & 1D \\
Parameter survey ($\omega_{c,i0}/q$) & 0.5 & 0.05 & 1 & 0.408 & 6400 & 1200 & 10 & 65536 & 0 & 0 & 1D \\
Parameter survey ($\omega_{c,i0}/q$) & 0.5 & 0.05 & 1 & 0.408 & 12800 & 1200 & 10 & 65536 & 0 & 0 & 1D \\
\enddata
\tablecomments{Simulation input parameters. The parenthetical labels in the Purpose column identify the control parameter(s) explored as part of the parameter survey. The remaining column definitions are given in Appendix \ref{appendix:simulation_parameters}.}
\end{deluxetable*}

\bibliography{main}
\bibliographystyle{aasjournalv7}

\end{document}